\UseRawInputEncoding
\documentclass[%
 aip,
 amsmath,amssymb,
 reprint,%
]{revtex4-1}

\usepackage{xr} 
\usepackage{graphicx}
\usepackage{dcolumn}
\usepackage{bm}

\usepackage[utf8]{inputenc}
\usepackage[T1]{fontenc}
\usepackage{mathptmx}
\usepackage{etoolbox}
\usepackage{mathptmx}
\usepackage{xcolor}
\usepackage{physics}
\usepackage[version=4]{mhchem}
\usepackage{xparse}

\newcommand{\corr}[1]{{\textcolor{black}{#1}}}

\makeatletter
\def\@email#1#2{%
 \endgroup
 \patchcmd{\titleblock@produce}
  {\frontmatter@RRAPformat}
  {\frontmatter@RRAPformat{\produce@RRAP{*#1\href{mailto:#2}{#2}}}\frontmatter@RRAPformat}
  {}{}
}%
\makeatother

\newcommand{\ktcf}[1]{\ensuremath{K_{#1}}}
\newcommand{\kB}{\ensuremath{k_{\mathrm{B}}}}
\newcommand{\rpH}[1]{\ensuremath{\mathcal{H}_{#1}}}
\newcommand{\rpV}[1]{\ensuremath{\widetilde{V}_{#1}}}
\NewDocumentCommand{\nmv}{m o}{%
  \vb{\tilde{#1}}%
  \IfValueT{#2}{_{(#2)}}%
}
\NewDocumentCommand{\nm}{m o}{%
  \tilde{#1}%
  \IfValueT{#2}{_{(#2)}}%
}
\newcommand{\rpqc}{\ensuremath{\vb{q}_{c}}}
\newcommand{\rppc}{\ensuremath{\vb{p}_{c}}}
\newcommand{\pmf}{\ensuremath{\bar{V}}}
\newcommand{\rcross}{\ensuremath{r_{x}}}
\newcommand{\rcentroid}{\ensuremath{r_{c}}}
\newcommand{\prob}{\ensuremath{\mathcal{P}}}

\newcommand{\eqn}[1]{Eq.~\eqref{eq:#1}}
\newcommand{\eqnn}[2]{Eqs.~\eqref{eq:#1} and~\eqref{eq:#2}}

\begin{document}

\title{Vibrational Spectra of Materials and Molecules from Partially-Adiabatic Elevated-Temperature Centroid Molecular Dynamics}

\author{Jorge Castro}
\affiliation{Max Planck Institute for the Structure and Dynamics of Matter, Hamburg, Germany}

\author{George Trenins}
\affiliation{Max Planck Institute for the Structure and Dynamics of Matter, Hamburg, Germany}

\author{Venkat Kapil}
\affiliation{Department of Physics and Astronomy, University College London, 7-19 Gordon
St, London WC1H 0AH, UK}
\affiliation{Thomas Young Centre and London Centre for Nanotechnology, 9 Gordon St, London WC1H 0AH}

\author{Mariana Rossi}
\affiliation{Max Planck Institute for the Structure and Dynamics of Matter, Hamburg, Germany}
 \email{mariana.rossi@mpsd.mpg.de}

\date{\today}

\begin{abstract}
Centroid molecular dynamics (CMD) incorporates nuclear quantum statistics into the calculation of vibrational spectra. However, when performed in Cartesian coordinates, CMD shows unphysical artifacts in certain vibrational bands, known as the curvature problem. Recent work showed that CMD spectra can be freed from the curvature problem by evolving the ring-polymer centroid on a potential of mean force (PMF) calculated at an elevated temperature ($T_e$-CMD). Here we present a partially-adiabatic implementation of $T_e$-CMD (PA-$T_e$-CMD), which eliminates the need for precomputed PMFs and instead yields the centroid force \textit{on the fly}. We introduce a two-temperature path-integral Langevin thermostat to achieve a temperature separation between centroid and internal modes of the ring polymer. Because it is paramount that the elevated temperature be chosen as low as possible for a given physical temperature in this formulation, we present a general scheme for its determination.
We benchmark PA-$T_e$-CMD against exact vibrational spectra for the isolated water monomer and discuss its performance for challenging anharmonic systems: the carbonic acid fluoride molecule (CAF) and the methylammonium lead iodide perovskite (MAPI). We conclude that PA-$T_e$-CMD mitigates the curvature problem and the steep increase in computational cost with decreasing temperature of conventional path-integral methods. We observe energy leakage from the hot internal modes to high-frequency centroid modes in some cases, \corr{which, nevertheless, only compromises the spectral lineshapes at lower  temperatures.}
While \corr{an adiabatic} setup based on a coarse-grained centroid PMF is still preferable when a good pre-trained PMF can be easily obtained,  PA-$T_e$-CMD presents a low-barrier single-shot setup for any system.
\end{abstract}

\maketitle

\section{\label{sec:Introduction} Introduction} 
 
 Nuclear quantum effects (NQEs) can be critical in vibrational spectroscopy, where they modulate the position, shape, and intensity of spectral features. Striking examples include redshifts and line-shape variations observed in hydrogen-bonded systems~\cite{Ikeda2021,Ikeda2023, Litman2019,LitmanFaradayDiscuss,Mengxu2025,Marsalek2017}, at aqueous interfaces~\cite{Rossi2021,Shepherd2023,Paesani2025}, in molecular crystals and in confined environments~\cite{Kapil2024, Rashmi2025}. Despite the theoretical foundation that quantum mechanics provides, solving the many-body time-dependent Schr\"odinger equation (TDSE) is computationally prohibitive for systems with more than a few tens of atoms. As a result, several approximate quantum many-body methods have been developed since the late 1990s, achieving varying degrees of success in bridging the gap between classical and quantum descriptions~\cite{JARRELL1996, Makri2003, Poulsen2003, WangThoss2003, Wang2015}.

Computational methods for vibrational spectroscopy that go beyond the standard harmonic treatment of nuclear motion broadly fall into two categories: methods that target a solution of the TDSE with varying levels of approximations and methods that try to approach quantum mechanics from modifications to classical dynamics. Techniques in the first category can provide near-exact quantum results. While these methods are well suited for small molecules or systems in which only a limited subset of degrees of freedom is relevant, their computational cost limits their applicability to larger and more complex systems. Prominent examples include the multi-configurational time-dependent Hartree method~\cite{Meyer1990}, quantum-classical path integrals~\cite{Makri2015Review}, hierarchical equations of motion~\cite{Tanimura1990}, and vibrational self-consistent field approaches~\cite{Bowman1978,Yagi2000,Kongsted2006,Bowman2008}.

In contrast, techniques in the second category are computationally efficient and well-suited for simulating all degrees of freedom in large systems. For example, simply using classical molecular dynamics to approximate vibrational motion is straightforward, but inherently ignores any NQEs. Approximations based on imaginary-time path integral molecular dynamics (PIMD) \corr{for distinguishable particles} provide an exact quantum statistical framework where time-independent observables can be sampled with classical trajectories on an extended phase-space spanned by a ring-polymer representation of the system~\cite{feynman2010quantum, ParrinelloRahman1998, ChandlerWolynes1981}. 
Extending this framework to simulate time-dependent properties, such as vibrational spectra, without resorting to computationally prohibitive real-time path integrals~\cite{BonellaCiccotti2010, MuhlbacherRabani2008, IlkMakri1994}, %
necessarily relies on (uncontrolled) approximations for the dynamics. Among the most widely used approaches are centroid molecular dynamics (CMD) \cite{Cao1994I, Jang1999} and (thermostatted) ring-polymer molecular dynamics [(T)RPMD] \cite{Craig2004, braams2006short, Habershon2013, Rossi2014, Rossi2018}. 

These methods approximate thermal quantum time-correlation functions in different ways.
CMD propagates dynamics on the centroid PMF generated by thermal fluctuations of the ring-polymer internal modes. In contrast, (T)RPMD considers the classical dynamics of the ring polymer in its full extended phase space. TRPMD is often favored for its computational efficiency, whereas CMD provides a more direct statistical mechanical interpretation through the centroid probability density. 
Although originally introduced heuristically, these methods were shown to be different approximations to Matsubara dynamics~\cite{Althorpe2021, WillattThesis2017}, thus representing different ways to join quantum statistics with classical dynamics.
Over the past decade, extensive work has demonstrated both the strengths and limitations of CMD and TRPMD for vibrational spectroscopy in a wide variety of systems spanning molecules, liquids and solids~\cite{Habershon2008, Marsalek2017, Rossi2014-Communication,Medders2015,Moberg2017,Reddy2017,Druzbicki2018,Markland2018,Hunter2018,Tong2020,Ikeda2021,Inakollu2021,Rossi2021,Kapil2022,Li2022,Lieberherr2023,Khuu2023,Ikeda2023,Shepherd2023,Bowles2024,Beckmann2024,Gomez2024,Nandi2025,Mengxu2025,Miranda2025,Rana2025,Litman2019, LitmanFaradayDiscuss, Trenins2019}.

A limitation of particular relevance to this work regards the CMD method. The well-documented ``curvature problem''~\cite{Witt, Ivanov2010} is an artifact that arises when CMD is performed in Cartesian coordinates~\cite{Trenins2019}. For systems where curvilinear vibrational motion is possible, coupling between stretching and angular modes of the ring-polymer can cause the centroid to drift outside the ring-polymer hull as the temperature is lowered. The result is an unphysical and pronounced red-shift and broadening of vibrational peaks~\cite{Witt, Ivanov2010, Rossi2014,Trenins2019}. In addition, the curvature problem can also distort vibrational intensities by altering the distribution of spectral weight. An illustration of this effect \corr{on a water monomer} is presented in \corr{Sec.}~S1 of the \corr{Supporting Information (SI)}.

Techniques such as quasi-centroid molecular dynamics (QCMD)~\cite{Trenins2019} and fast-QCMD~\cite{Fletcher2020} avoid the curvature problem by evolving the system on the PMF of a carefully chosen curvilinear function of ring-polymer bead positions.
However, extending both QCMD and fast-QCMD to general systems requires a universal free-energy fitting procedure and careful selection of curvilinear coordinates~\cite{Haggard2021} to ensure accurate spectra and comprehensive configurational sampling.

An alternative method that mitigates the curvature problem  is the ``elevated-temperature'' CMD method~\cite{Musil2022} ($T_e$-CMD). In this approach, the centroid PMF is computed at an elevated temperature \( T_e \), where the curvature problem is negligible, while the system evolves at the physical temperature. In the original implementation \cite{Musil2022}, a path-integral coarse-graining force-matching protocol was used in order to train a machine-learning force-field for the centroid PMF based on a reference PIMD simulation. This approach is referred to as PIGS. To attenuate the curvature problem, the reference PIMD simulations are performed at a high temperature $T_e$. After the PMF is trained, the evaluation of a $T_e$-CMD vibrational spectrum only requires running classical molecular dynamics at any desired physical temperature and constructing the desired correlation function. The resulting spectra are free of the curvature problem and still capture the relevant quantum statistical effects related to zero-point energy motion. \( T_e \)-PIGS was shown to be successful for  molecular and condensed-phase systems, even for cryogenic physical temperatures~\cite{Musil2022, Kapil2024, Zaporozhets2024}.

Performing a $T_e$-PIGS simulation inherently involves two steps: training the PMF and subsequently running the simulation. Because the PMF is typically constructed by combining a baseline \corr{potential energy surface} with an ML model that upgrades it to the PMF, it also opens the possibility of \corr{fitting errors.} 
However, just as in standard CMD, it should also be possible to calculate vibrational spectra within the $T_e$-CMD framework with an on-the-fly partially adiabatic (PA) protocol. In this work, we introduce precisely this approach. In partially-adiabatic $T_e$-CMD (PA-$T_e$-CMD) the \corr{forces associated with the} centroid PMF is computed during the simulation, eliminating the need for precomputed models. The method combines the mass rescaling strategy of PA-CMD~\cite{Hone2006}, which assigns lighter fictitious masses to the internal ring-polymer modes to accelerate their dynamics, with the elevated-temperature \textit{ansatz} enforced through a two-temperature path-integral Langevin thermostat~\cite{Ceriotti2010}. This setup efficiently samples the high-temperature PMF while keeping the centroid at the physical temperature. As a result, PA-\( T_e \)-CMD offers a single-shot alternative to $T_e$-PIGS, improving transferability and ease of use while preserving the core features of the elevated-temperature approach. \corr{The main trade-off} is the need of using smaller time-steps in the simulation. 

In the following, we describe the PA-$T_e$-CMD method in detail, provide guidelines for choosing $T_e$ for different systems, and benchmark its performance against exact spectra, where available, and against other path-integral-based methods such as CMD, TRPMD, and $T_e$-PIGS. For an isolated water monomer, PA-$T_e$-CMD closely reproduces the results of $T_e$-PIGS and the exact vibrational density of states and infrared spectra. For methylammonium lead iodide (MAPI), accurate spectra for its tetragonal and orthorhombic phases are obtained from PA-\( T_e \)-CMD and compare well with $T_e$-PIGS, showing that the latter performs well even when the lower temperature simulations are run with a PMF trained on the high-temperature cubic phase.  Finally, for the carbonic acid fluoride molecule [H$_2$CO$_3$F]$^-$ (CAF) we find it more difficult to obtain accurate centroid PMFs \corr{using the training-data acquisition strategy proposed for} $T_e$-PIGS \corr{in the original work~\cite{Musil2022}, due to the strong temperature dependence of chemical bonding in the weakly bound complex.} In this case, PA-$T_e$-CMD offers a good alternative for obtaining accurate vibrational fingerprints. Overall, we observe that it is more challenging to 
maintain the temperature separation between centroid and internal modes of the ring polymer in PA-$T_e$-CMD when the physical temperature is very low. Nevertheless, PA-$T_e$-CMD proves to be simple to set up for any system and to provide a reliable route to obtain vibrational spectra of condensed-phase systems and molecules, in regimes where nuclear quantum coherences are naturally \corr{quenched} by temperature and many-body interactions.

\section{\label{sec:theory} Methods}

\subsection{Centroid Molecular Dynamics}
\label{CMD}

CMD provides an approximation to quantum Kubo-transformed time-correlation functions (KTCF)~\cite{Cao1994I, Cao1994II, Jang1999, Hone2006} of the form
\begin{equation} 
\begin{aligned} 
\ktcf{AB}(t,\beta) = \frac{1}{\beta Z} \int_0^{\beta} d\lambda \ \text{tr}[e^{-(\beta-\lambda)\hat{H}}\hat{A}e^{-\lambda\hat{H}}\hat{B}(t)],
\end{aligned} 
\end{equation}
where $Z=\text{tr}[e^{-\beta\hat{H}}]$ is the quantum canonical partition function, $\beta = 1/\kB T$ is the inverse temperature, $\hat{B}(t)=e^{+i\hat{H}t/\hbar}\hat{B}e^{-i\hat{H}t/\hbar}$, $\hat{H}$ is the Hamiltonian of the system and $t$ is the physical time. 

Like other PIMD-based methods, CMD can be obtained from the dynamics of a fictitious ring-polymer Hamiltonian. For a one-dimensional system with $P$ replicas, this Hamiltonian can be written in terms of the free ring-polymer normal modes as
\begin{equation} 
\begin{aligned} 
\rpH{P} = \sum_{k=0}^{P-1} \left( \frac{\nmv{p}[k]^{2}}{2m} + \frac{1}{2} m \omega_k^2 \nmv{q}[k]^{2} \right) + \rpV{P}(\nmv{q}), \label{eq:rp-ham-nm}
\end{aligned} 
\end{equation}
where $\nmv{q}$ and $\nmv{p}$ are the positions and momenta, respectively, transformed into the free ring-polymer normal-mode basis. For the transformation, we employ the same normalization convention as Ref.~\citenum{TreninsJCP2018}, such that the zeroth ring-polymer normal mode is the centroid $\rpqc \equiv \nmv{q}[0]$. Here, $\omega_k = 2P/(\beta \hbar)  \sin(k\pi/P)$ are the frequencies of the normal modes, and $\rpV{P}(\nmv{q})$ is the average of the physical potential over the ring-polymer beads. For multidimensional systems, this expression generalizes straightforwardly by summing over all Cartesian degrees of freedom.

For operators $\hat{A}$ and $\hat{B}$ that are either functions of the position operator or proportional to the momentum operator, the CMD approximation to the KTCF can then be written as

\begin{equation} 
\label{eq:Kubo}
\begin{aligned}
    \ktcf{AB}(t, \beta) & \approx \frac{1}{Z} \int d \rpqc \int d \rppc \\ & {} \times A[\rpqc(0)] B[\rpqc(t)] e^{-\beta \left( \frac{\rppc^2}{2m} + \pmf(\rpqc,\beta) \right)},
\end{aligned}
\end{equation}
where $\pmf(\rpqc, \beta)$ is the PMF for the centroid, computed by averaging over the non-centroid normal modes of the ring polymer
\begin{equation} 
\begin{aligned}
\pmf(\rpqc) = -\frac{1}{\beta} \ln [\lim_{P \rightarrow \infty}\langle \delta(\nmv{q}[0] - \rpqc)\rangle]. \label{eq:centroid-pmf}
\end{aligned}
\end{equation}
The thermodynamic average $\langle \cdot \rangle$ is calculated with the ring-polymer Hamiltonian of Eq.~\eqref{eq:rp-ham-nm} \corr{in the canonical ensemble}. As an example, to obtain  the vibrational density of states, Eq.~\eqref{eq:Kubo} needs to be evaluated with $A$ and $B$ set to the centroid velocity.

The dynamics \corr{of the centroid follows simple} classical equations of motion\corr{:}
\begin{equation}
\begin{aligned}
\dv{\rpqc}{t} & = \frac{\rppc}{m}, \\
\dv{\rppc}{t} & = - \pdv{\pmf(\rpqc)}{\rpqc}.
\end{aligned}
\end{equation}

\corr{To propagate these equations,} the centroid PMF can either be precomputed (e.g., on a grid or with a fitted functional form) or \corr{its negative gradient can be} computed on the fly. The latter is achieved by imposing an adiabatic separation between the dynamics of the centroid and non-centroid modes. In partially adiabatic \corr{(PA-)} CMD \cite{Hone2006}, the fictitious masses of the non-centroid normal modes are rescaled to a fraction of the value of the lightest  chemical species in the system, while keeping the mass of the centroids equal to the physical mass. The PA-CMD Hamiltonian can be written as
\begin{align} 
\rpH{P} & = \sum_{k=0}^{P-1} \left( \frac{\nmv{p}[k]^2}{2m_k} + \frac{1}{2} m_k \Omega^2 \nmv{q}[k]^{2} \right) + \rpV{P}(\nmv{q}), 
\end{align} 
where $\Omega$ is the frequency of adiabatic separation,  $m_k = m / \sigma_k^2$, and $\sigma_k$ is the adiabatic scaling parameter,
\begin{equation}
    \label{eq:ad-scaling}
    \sigma_k = \begin{cases}
        1 & \qqtext*{for} k = 0, \\
        \Omega / \omega_k & \text{otherwise}.
    \end{cases}
\end{equation}
The scaling is defined such that $m_k \Omega^2 = m \omega_k^2$, so the ring-polymer quantum statistics are not altered by this renormalization.

The choice of $\Omega$ is critical to the accuracy and efficiency of PA-CMD. By setting this frequency to a value much larger than any physical vibration in the system, one enables dynamical sampling of $\pmf(\rpqc)$ as the centroid moves on a slower timescale. However, a larger $\Omega$ increases the highest dynamical frequency in the system, requiring smaller time steps for accurate integration of the equations of motion. In practice, $\Omega$ is selected as a compromise between convergence and computational efficiency, balancing accurate PMF sampling with a reasonably large timestep. Efficient sampling of the PMF requires that strongly coupled thermostats are applied to the non-centroid modes, and we note that the choice of thermostat and its parameters also affects the choice of $\Omega$, as discussed in Ref.~\citenum{Rossi2014-Communication}.

\subsection{The Elevated Temperature Ansatz%
\label{sec:ansatz}}

It is well documented that at low temperatures, a vibrational spectrum calculated from the CMD correlation function in Eq.~\eqref{eq:Kubo} presents the so-called curvature problem~\cite{Ivanov2010}. Ultimately, this problem stems from the use of Cartesian coordinates to calculate the PMF~\cite{Trenins2019}, and different techniques have been proposed to mitigate it. In particular, Musil \textit{et al.}~\cite{Musil2022} introduced the elevated temperature ($T_e$) \textit{ansatz} that effectively mitigates the curvature problem because at $T_e$ a curvilinear centroid is well approximated by its Cartesian counterpart. 

This approach exploits the fact that in a system in the vibrational ground state ($\hbar \omega \gg \kB  T$), the exact quantum time-correlation function is temperature-independent.
Hence, a low-temperature KTCF can be obtained by simulating the system at an elevated temperature $T_e$ that is high enough to suppress the curvature problem but low enough for the system to remain in the vibrational ground state.
In this limit, the Fourier transform of the KTCF satisfies\cite{Musil2022}
\begin{equation}
   \ktcf{AB}(\omega,\beta) = \frac{\beta_{e}}{\beta}\,\ktcf{AB}(\omega,\beta_{e}), \label{eq:high-t}
\end{equation}
where $\beta_e = 1/(k_BT_e)$. The \corr{right hand side of the equation is then approximated to a form which is parametric to the elevated temperature $T_e$ with}
\begin{equation}
\begin{aligned}
  \ktcf{AB}(\omega,\beta) \approx K^{T_{e}}_{AB}(\omega,\beta; \beta_{e}) = \frac{1}{Z} \mathfrak{F} \bigg[ 
    \int d\mathbf{q}_{c} \, d\mathbf{p}_{c} \,
    A[\mathbf{q}_{c}(0)]\, B[\mathbf{q}_{c}(t)]\\ 
    \times \exp\Bigl(-\beta \Bigl[\frac{\mathbf{p}_{c}^2}{2m} + \bar{V}(\mathbf{q}_{c}, \beta_{e})\Bigr]\Bigr)
    \bigg], \label{eq:kubo-w-hight}
\end{aligned}
\end{equation}
which is exact in the harmonic and in the classical limit. \corr{We call this approximation $T_e$-CMD in the following.}
\corr{The elevated temperature \emph{ansatz} requires microcanonical centroid dynamics on the PMF estimated at \(T_e\), while sampling initial conditions at the physical temperature $T$.} This has proven to be an excellent approximation for vibrational spectra \cite{Musil2022,Ravindra2024,Ravindra2024NuclearQE,Paesani2025,Richa2025}.

\subsection{A Practical Recipe for Choosing the Elevated Temperature \label{subsec:choosing-te}}

The elevated temperature for $T_e$-CMD should be chosen with care. It must be high enough to avoid the curvature problem, but not so high that it breaks the approximation in Eq.~\eqref{eq:kubo-w-hight}. In Ref.~\citenum{Musil2022}, it was proposed to determine $T_e$ by performing PIMD simulations at a range of temperatures and selecting the lowest temperature at which the mode of the radial distribution of centroid positions aligns with the mode of the exact (physical) quantum radial distribution along that coordinate.

Here we present an alternative procedure, that does not require running PIMD simulations. As discussed by Trenins and Althorpe in Ref.~\citenum{TreninsJCP2018}, for vibrational modes coupled to curvilinear motion, there is a temperature below which artificial instantons
are likely to emerge for centroid-constrained ring-polymer configurations sampled by CMD. In order to determine this temperature, the authors of Ref.~\citenum{TreninsJCP2018} have conveniently defined a crossover radius $\rcross$. If \corr{the length of a centroid bond vector, $r = \abs*{\nmv{q}[0]}$, lies below $\rcross$, that bond forms artificial instantons, and both the shape and position of the corresponding} vibrational peak are dramatically altered. There is a simple and general expression for the $\rcross$ of a single vibrational coordinate \corr{coupled to free rotation, obtained by} minimizing the ring-polymer energy subject to the centroid constraint. This expression reads
\begin{equation}
\rcross(\beta) = - \frac{\beta}{2 \mu \pi} \frac{\partial V(r-\corr{r_{\text{eq}}})}{\partial r} \Bigg\vert_{r=\rcross},
\label{eq:rcross}
\end{equation}
where $\mu$ is the (reduced) mass of the vibrational motion, $V$ is the radial part of the potential along that coordinate and $\corr{r_{\text{eq}}}$ is the equilibrium position. 

For our purposes, it is convenient to define three temperature regimes, 
according to how much of the centroid thermal distribution extends beyond 
$\rcross$. The high-temperature regime ($T > T_{\text{high}}$) corresponds to a distribution with a negligible weight beyond $\rcross$, as in the top panel of Fig.~3 in Ref.~\citenum{TreninsJCP2018}, which we reproduce in \corr{Fig.~S3 of the SI}. Here, the curvature problem is negligible and the elevated-temperature \emph{ansatz} is dropped, so that the ``elevated'' and the physical temperatures can be kept equal, $T_e = T_{\text{phys}}$. 

In the low-temperature regime ($T < T_{\text{low}}$), $\rcross$ substantially overlaps with the centroid distribution, as in the bottom panel of Fig.~S3. This is linked to line shape deformation in standard PA-CMD and a growing deviation between the distribution modes for centroid and \emph{quasi}centroid coordinates (the compact curvilinear coordinates introduced in Ref.~\citenum{Trenins2019} -- also see Sec.~\ref{sec:Introduction}). In this regime, the curvature problem affects the centroid PMF even around equilibrium and should not be used for vibrational dynamics simulations. We, therefore, impose a lower bound on the elevated temperature, $T_e \geq T_{\text{low}}$.

In the intermediate temperature range, $T_{\text{low}} < T < T_{\text{high}}$, $\rcross$ overlaps a non-negligible part of the centroid distribution (middle panel of Fig.~S3), but is still far enough not to corrupt the PMF around the equilibrium geometry. When the physical temperature drops below $T_{\text{high}}$ we propose to pick an elevated temperature $T_e > T_{\text{phys}}$ from this range. $T_e$ should be as low as possible to fulfill the low-temperature criterion of Sec.~\ref{sec:ansatz} but high enough for $\rcross$ to shift far away from the mean of the 
centroid thermal distribution $\prob(\rcentroid, \beta_{\text{phys}})$ computed for the PMF $\pmf(\mathbf{q}_{c}, \beta_{e})$.

\begin{figure}[ht]
    \centering
    \includegraphics{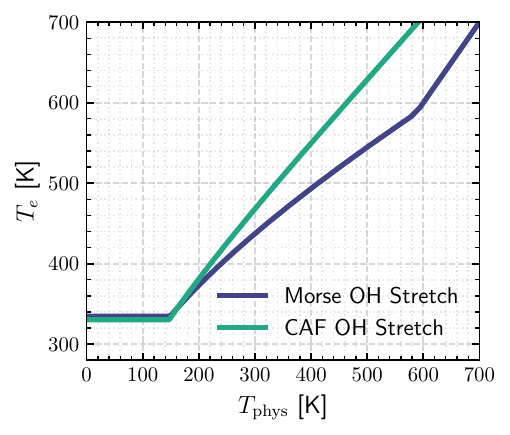}
    \caption{Elevated temperature $T_e$ as a function of the physical temperature $T_{\mathrm{phys}}$ for the analytical radial OH Morse potential with the same parameters as in Ref.~\citenum{Trenins2019} (blue) \corr{and for a numerically evaluated O–H bond-stretching potential of the carbonic acid fluoride (CAF) molecule (green). The CAF potential was obtained by displacing the hydrogen atom along the O–H bond direction.}}
    \label{fig:te-recipe}
\end{figure}

In polar coordinates, we are not aware of an analytical closed-form solution for the probability distribution of $\rcentroid$ on the centroid PMF of Eq.~\eqref{eq:centroid-pmf}, even when the potential is harmonic in the radial coordinate. However, following from the argument in Sec.~\ref{sec:ansatz}, at sufficiently low temperatures such that the system still remains at the vibrational ground state, but high enough that the ring polymer is not overly spread around the curviliner potential energy profile, we can approximate the PMF as harmonic along this coordinate, leading to a linear force on $r_c$ and a Gaussian thermal distribution. In other words, we consider
\begin{equation}
\prob(\rcentroid, \beta) \propto \exp\left[- \beta \mu \omega_{\mathrm{harm}}^2 (\rcentroid-\corr{r_{\text{eq}}})^2\right], \label{eq:distribution}
\end{equation}
where $\mu \omega_{\mathrm{harm}}^2 = \partial^2 V(r-\corr{r_{\text{eq}}})/\partial r^2|_{r=\corr{r_{\text{eq}}}}$. The standard deviation is 
\begin{equation}
    \sigma(\beta) = (\beta \mu \omega_{\mathrm{harm}}^2)^{-1/2}.%
    \label{eq:rc-sdev}
\end{equation} 
More controlled approximations to this distribution can be considered in future work, e.g., based on steepest-descent integration of Eq.~\eqref{eq:centroid-pmf}. This kind of analysis would account for the temperature-dependent shift in the mode of the distribution, which, in reality, is not strictly fixed at $\corr{r_{\text{eq}}}$, as well as for anharmonicity along the radial coordinate. The quality of the approximation we use indeed becomes less justified at very low temperatures. However, the current treatment is sufficient to get a robust estimate for an optimal $T_e$ and has the added benefit of simplicity. 

For our recipe, we numerically solve Eq.~\eqref{eq:rcross} for a given anharmonic $V(\rcentroid)$ and define
\begin{subequations}
\label{eq:T-bounds}
\begin{align}
\label{eq:T-high}
T_{\text{high}} & = 1/\kB \beta_{\text{high}} & & \text{s.t.} & \corr{r_{\text{eq}}} - r_x({\beta_{\text{high}}}) & = 6 \sigma(\beta_{\text{high}}), \\
\label{eq:T-low}
T_{\text{low}} & = 1/\kB \beta_{\text{low}} &  & \text{s.t.} & \corr{r_{\text{eq}}} - r_x({\beta_{\text{low}}}) & = 4 \sigma(\beta_{\text{low}}).
\end{align}
\end{subequations}
The choice of multiples of $\sigma$ in Eq.~\eqref{eq:T-bounds} is somewhat arbitrary but was found to give good results in our numerical tests. 

To find the optimal elevated temperature in all regimes, we then invert the numerical solutions to Eq.~\eqref{eq:rcross} to obtain the crossover temperature for a given centroid radial displacement $\beta_x(\rcentroid)$. For any physical temperature $T_{\text{phys}} < T_{\text{high}}$, we compute 
\corr{\begin{equation}
\beta_{\mathrm{candidate}} = \beta_x[\corr{r_{\text{eq}}} - 6\sigma(\beta_{\text{phys}})].\label{eq:tcandidate}
\end{equation}
The} elevated temperature $T_e$ is then set to the largest of  $T_{\mathrm{candidate}}$ and $T_{\text{low}}$. This choice is motivated by the fact that the approximation in Eq.~\eqref{eq:distribution} becomes worse at very low temperatures.  Above $T_{\text{high}}$, we obviously set $T_e = T_{\text{high}}$. The procedure only requires the potential-energy profiles \corr{along the bonds that are prone} to the curvature problem and, thus, gives a simple and general prescription for $T_e$.

\corr{We now summarize the procedure for selecting the elevated temperature $T_e$, illustrated in the flowchart in Fig.~S4 of the SI. We identify the bonds that are most susceptible to the curvature problem. Typically, these are covalent bonds between hydrogen and a heavier atom (e.g., \ce{O-H}, \ce{N-H}, \ce{C-H}), which are involved in high-frequency vibrational modes and couple to low-frequency librations or rotations. For each selected bond, a one-dimensional potential energy curve is mapped by displacing the light atom along the bond direction in small increments around its equilibrium geometry and computing single-point energies for each configuration. The reduced mass for each stretching coordinate is estimated from the atomic masses of the two bonded atoms. We then apply the following procedure:
\begin{itemize}
\item From these potentials, $r_x(\beta)$ is obtained using \eqn{rcross}, and $\sigma(\beta)$ is obtained using \eqn{rc-sdev}.
\item We calculate $T_{\text{low}}$ and $T_{\text{high}}$ through \eqnn{T-high}{T-low}, respectively.
\item For a given $T_{\text{phys}}$ we apply:
\begin{itemize}
    \item If $T_{\text{phys}} < T_{\text{high}}$, we compute Eq.~\eqref{eq:tcandidate}. We choose $T_e$ as the largest of $T_{\mathrm{candidate}}$ and $T_{\text{low}}$.
    \item If $T_{\text{phys}} \geq T_{\text{high}}$, $T_e = T_{\text{high}}$
\end{itemize}
\end{itemize}
For systems with multiple bonds susceptible to the curvature problem, we select the highest among the $T_e$ values estimated for the individual bonds.
For illustration, the procedure is shown for the \ce{C-H} and \ce{N-H} stretching coordinates of MAPI in \corr{Sec.}~S4 and \corr{Fig.}~S5 of the SI.}

For the 2D champagne-bottle potential widely studied in this context\cite{Ivanov2010, Rossi2014, TreninsJCP2018, Musil2022, Trenins2019}, the solutions for $T_e$ at different physical temperatures $T_{\mathrm{phys}}$ are shown in Fig.~\ref{fig:te-recipe}. \corr{In the same figure, we also show the solution for the O–H bond-stretching coordinate of the carbonic acid fluoride (CAF) molecule, obtained by displacing the hydrogen atom along the O–H bond direction},  as an example of an anharmonic vibrational motion where the curvature problem is very pronounced. We observe 
that our procedure allows us to select
$T_e < T_{\mathrm{high}}$ tailored to specific physical temperatures.  
Numerical tests on the water monomer, CAF, and MAPI presented in \corr{Sec.}~S5 of the SI validate this procedure.

\subsection{Implementing Partially-Adiabatic CMD with the Elevated Temperature Ansatz}
\label{subsec:implementation}

The original implementation of elevated-temperature CMD is called \(T_e\)-PIGS.\cite{Musil2022} In that approach, the centroid force is extracted from high-temperature path-integral simulations, and a machine-learned interatomic potential (MLIP) is subsequently trained to approximate the PMF at \(T_e\) by using path-integral coarse-graining techniques. \(T_e\)-PIGS offers accuracy and efficiency advantages. The dynamics are accurate (within the approximations of the method) because there is a strict adiabatic separation of the centroid and the internal modes of the ring polymer and they are efficient because one simply performs classical molecular dynamics on the pre-trained centroid PMF. 

One of the main drawbacks is the setup overhead: the centroid PMF is a free-energy surface that must be parameterized separately, often by combining a baseline ML interatomic potential with an additional model, which introduces extra training steps, even  compared to using a standard MLIP.
This can make \( T_e \)-PIGS cumbersome for systems whose relevant free-energy landscape spans many distinct configurations that may differ considerably from those sampled at the elevated training temperature, and for systems where current ML potentials are not sufficiently accurate.

The partially adiabatic formulation of elevated-temperature CMD (PA-T$_e$-CMD) addresses these drawbacks. PA-T$_e$-CMD computes the centroid PMF on the fly, eliminating the need for precomputed PMFs. This is achieved by implementing a two-temperature path-integral Langevin equation thermostat (2T-PILE). The 2T-PILE thermostat extends the standard path-integral Langevin equation (PILE) thermostat to simultaneously thermalize the centroid at the target temperature \(T_{\mathrm{phys}}\) while equilibrating the internal modes at the elevated temperature \(T_e\).

The 2T-PILE formulation follows the general structure of PILE thermostats,\cite{Ceriotti2010} introducing stochastic and dissipative forces for each normal mode \(k\) of the ring polymer,
\begin{equation}
\nm{p}[k] \leftarrow c_{1,(k)} \nm{p}[k]  + \sqrt{\frac{m}{\beta_{(k)}}}\, c_{2,(k)} \xi_{(k)},
\label{eq:PILE}
\end{equation}
where \(\xi_{(k)}\) is Gaussian white noise with zero mean and unit variance. The coefficients \(c_{1,(k)}\) and \(c_{2,(k)}\) are defined as:
\begin{align}
c_{1,(k)} &= \exp\!\left[-\frac{\Delta t}{2}\,\gamma_{(k)}\right], \quad
c_{2,(k)} = \sqrt{1 - \left(c_{1,(k)}\right)^{2}}.
\end{align}
The friction coefficients are defined by
\begin{align}
\gamma_{(k)} =
\begin{cases}
\dfrac{1}{\tau_{0}}, & k = 0, \\
\lambda\, 2\omega_{k} \sigma_{k}, & k > 0, \label{eq:pile-l}
\end{cases}
\end{align}
where \(\omega_{k}\) is the frequency of the ring-polymer mode \(k\) and $\sigma_k$ is the partially adiabatic scaling parameter in Eq.~\eqref{eq:ad-scaling}. 
The parameter \(\lambda\) controls the degree of damping of the internal ring-polymer modes by the Langevin thermostats, whereas \(\tau_0\) governs the relaxation time of the thermostat attached to the centroid. 

Compared to the standard setup, 2T-PILE only modifies the temperature entering Eq.~\eqref{eq:PILE}. It assigns distinct inverse temperatures \(\beta_{(k)}\) to normal modes with the index $k$,
\begin{align}
\beta_{(k)} =
\begin{cases}
\dfrac{1}{k_{B} T_{\mathrm{phys}}}, & k = 0, \\
\dfrac{1}{k_{B} T_{e}}, & k > 0,
\end{cases}
\end{align}
so that the adiabatically separated fluctuation modes sample the thermodynamic ensemble consistent with $T_e$, while the centroid evolves on the resulting PMF at a temperature $T_{\text{phys}}$. We stress that the ring-polymer springs are defined such that they are consistent with the elevated temperature as well, i.e., $\omega_k = 2P/(\beta_e \hbar)  \sin(k\pi/P)$.

\begin{figure}
    \centering
    \includegraphics{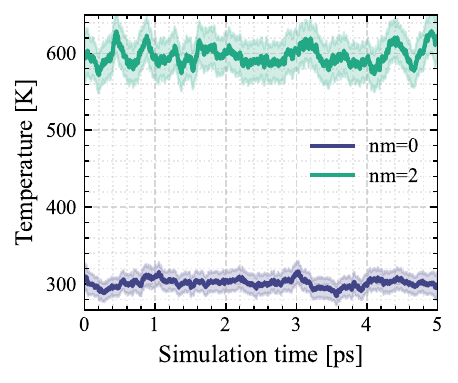}
    \caption{
    Time evolution of the temperatures of the centroid mode (nm $=0$, blue line) and the second internal normal mode of the ring polymer (nm $=2$, green line) during a PA-$T_e$-CMD simulation of an isolated water monomer at $T_{\mathrm{phys}} = 300$~K and $T_e = 600$~K. Solid lines show centered moving averages, and shaded areas indicate the 95\% confidence intervals computed from a moving window. Both modes exhibit stable and distinct temperatures throughout the simulation, as expected from the 2T-PILE thermostat.
    }
    \label{fig:2T-thermostat}
\end{figure}

As discussed in Ref.~\citenum{Rossi2014-Communication}, for PA-CMD protocols, the parameter $\lambda$ needs to remain in the under-damped regime to minimize artificial coupling between internal and centroid modes. 
Moreover, in usual PA-CMD protocols, one sets $\tau_0$ to be very large, effectively switching off the centroid thermostat. This is done to minimize the disturbance to the centroid dynamics, ensuring the correct evaluation of time-correlation functions according to Eq.~\eqref{eq:Kubo}.

In PA-\(T_e\)-CMD, we found that a weak global thermostat attached to the centroid helps prevent kinetic energy leakage from internal ring-polymer modes. To achieve this, instead of using a local Langevin thermostat attached to the centroid mode like in Eqs.~\eqref{eq:PILE} and \eqref{eq:pile-l}, we adopt the same strategy as in Ref.~\citenum{Ceriotti2010} and attach instead a ``global'' thermostat to the centroid, that acts on the kinetic energy of the system, instead of the individual momenta. This thermostat is the stochastic velocity rescaling (SVR) thermostat,\cite{Bussi2007} which only minimally disturbs dynamics if its relaxation time, which we will call $\tau$, is not too small.

We have implemented this thermostat in the i-PI program package.\cite{Litman2024}
\corr{Fig.}~\ref{fig:2T-thermostat} illustrates the action of the 2T-PILE thermostat with $T_e = 600$ K and $T_{\mathrm{phys}} = 300$ K for an isolated water monomer. 
It shows the successful independent thermalization of the centroid and non-centroid modes at distinct temperatures. \corr{An analogous analysis for the MAPI system is presented in \corr{Fig.}~S9 of the SI, which confirms that the thermostatting strategy is robust for systems with a large number of degrees of freedom.}
We benchmark the performance of different \(\lambda\) and \(\tau\) parameters in \corr{Sec.}~\ref{sec:result-bench}.

\begin{figure*}[ht]
    \centering
    \includegraphics[width=0.9\textwidth]{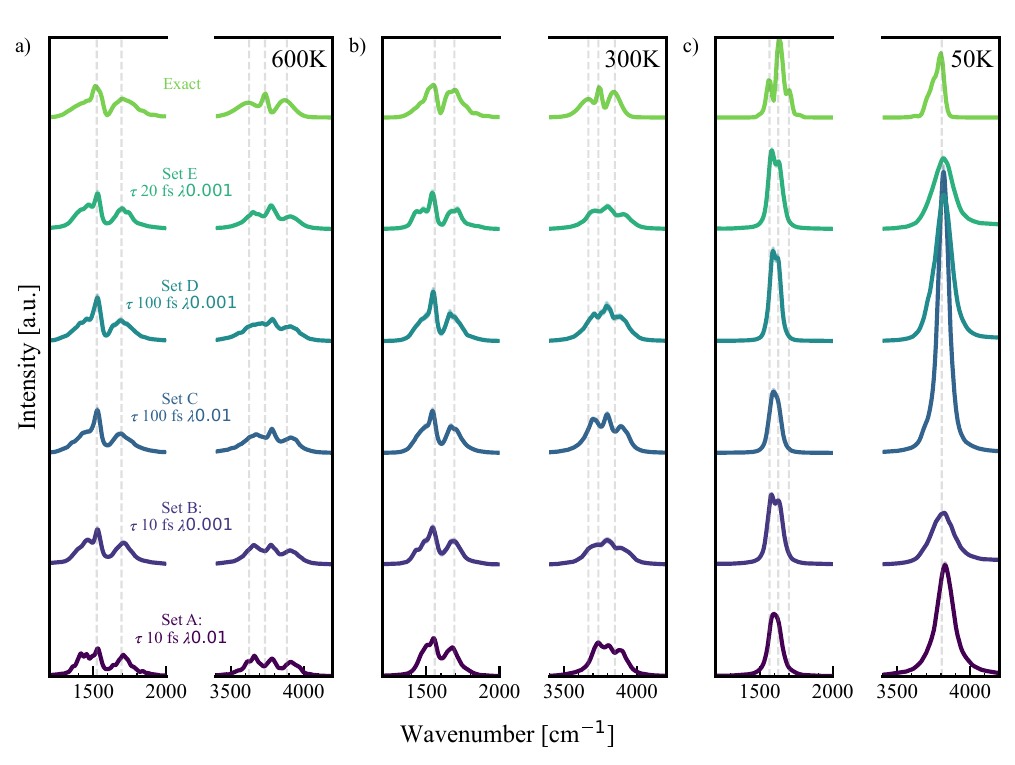}
    \caption{Bending (1100--2000 cm$^{-1}$) and O--H stretching (3500--4200 cm$^{-1}$) infra-red vibrational bands of a water monomer obtained via PA-T$_e$-CMD with five parameter sets compared to the DVR reference~\cite{Tennyson2004} at (a) 600~K, (b) 300~K, and (c) 50~K.     }
    \label{fig:benchmarking1}
\end{figure*}

\section{\label{sec:comp} Computational Details}

This section details the computational protocols employed in our PA-\( T_e \)-CMD and \( T_e \)-PIGS simulations for three representative systems: a water monomer, the carbonic acid fluoride (CAF) molecule and the methylammonium lead iodide (MAPI) perovskite. These systems were selected to assess the performance and limitations of our method across diverse vibrational environments. The water monomer serves as a benchmark due to its well-characterized vibrational spectrum and the availability of exact quantum reference data. CAF presents strong anharmonicity for some vibrational motions, related to the presence of a strong ionic hydrogen bond. Finally, MAPI presents mild anharmonicity, but exhibits multiple polymorphs across different temperatures.

We employed the following interatomic potentials for each system: (i) the Partridge--Schwenke potential~\cite{PS1997} for the water monomer; (ii)  a MACE~\cite{Batatia2022mace,Batatia2022Design} potential trained on density-functional theory data using the B3LYP functional~\cite{Becke1993,Stephens1994} and including pairwise dispersion interactions~\cite{Tkatchenko2009} for CAF ; and (iii) a MACE potential trained on density-functional theory data using the HSE06~\cite{Heyd2003,Krukau2006} hybrid range-separated functional and incorporating many-body dispersion interactions~\cite{Tkatchenko2012} for MAPI. Additional details on the data sets, the training procedure and the validation of the MACE potentials are provided in the Supporting Information Sec.S8.

All PA-$T_e$-CMD simulations were run with the i-PI code~\cite{Litman2024}. 
For comparison, vibrational spectra were also computed using classical molecular dynamics, TRPMD, PA-CMD and \( T_e \)-PIGS, using the i-PI code as well. The \( T_e \)-PIGS simulations were performed following the methodology of Ref.~\citenum{Kapil2024} and utilized the same baseline potentials as PA-\( T_e \)-CMD for each system. The \( T_e \)-PIGS centroid PMF was trained with \( T_e = 600 \)~K for the water monomer, for comparison with previous work \cite{Musil2022}. For MAPI, the PMF for \( T_e \)-PIGS was trained at \( T_e = 500 \)~K and at the cubic phase. Finally, for CAF we used PMFs trained at several temperatures, as discussed below. \corr{Further computational details on the elevated-temperature simulations, as well as on the additional simulation methods, are provided in Sections~S7 and~S9 of the SI.}

\section{Results}
\label{sec:results}
\subsection{Benchmarks on the Water Monomer}
\label{sec:result-bench}

\begin{figure*}[ht]
    \centering
    \includegraphics[width=0.9\textwidth]{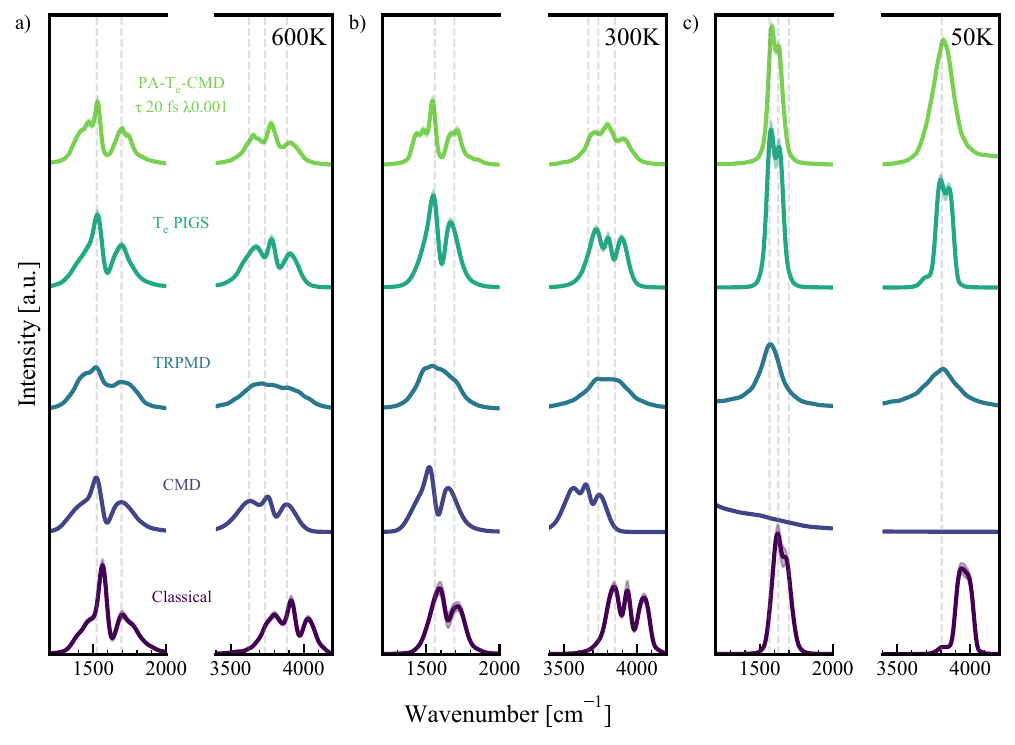}
    \caption{Comparison between PA-T$_e$-CMD (Set E), 
    T$_e$-PIGS, standard CMD, TRPMD, and classical MD against the DVR reference~\cite{Tennyson2004} for bending (1100--2000 cm$^{-1}$) and O--H stretching (3500--4200 cm$^{-1}$) IR vibrational bands of a water monomer at (a) 600~K, (b) 300~K, and (c) 50~K.} 
    \label{benchmarking2}
\end{figure*}

We start by benchmarking the PA-T$_e$-CMD method for the water monomer. Fig.~\ref{fig:benchmarking1} shows the O--H bending (1100--2000 cm$^{-1}$) and stretching (3500--4200 cm$^{-1}$) vibrational bands in the IR spectrum of the water monomer.  \corr{Our references are the exact spectra obtained from solving 
the Schr\"{o}dinger equation for the same potential, using a discrete variable representation (DVR)~\cite{Colbert1992} implemented in the DVR3D program suite by Tennyson et~al.\cite{Tennyson2004}} PA-T$_e$-CMD results are shown for different combinations of $\tau$ and $\lambda$ parameters, discussed in \corr{Sec.}~\ref{subsec:implementation}. 

We tested the parameter values $\tau = 10$, $20$, $100$~fs and $\lambda = 0.01$, $0.001$. The adiabatic separation frequency was fixed at \( \Omega = 24000 \)~cm\(^{-1} \). 
These values were chosen to explore the tradeoffs between thermostat strength, sampling efficiency, and spectral accuracy. Smaller $\tau$ values introduce larger disturbances to the dynamics of the centroid, but maintain its temperature more efficiently. \corr{Small $\lambda$ values take the fluctuation-mode dynamics to the underdamped regime and make on-the-fly PMF sampling less efficient, but also mitigate spurious dynamical coupling between the adiabatically separated fluctuation modes and the centroid.}

\corr{Fig.}~\ref{fig:benchmarking1}a–c compares PA-$T_e$-CMD spectra across five combinations of the parameters $\tau$ and $\lambda$, benchmarked against the exact DVR reference. For future reference, we label them as follows: Set A ($\tau = 10$~fs, $\lambda = 0.01$), Set B ($\tau = 10$~fs, $\lambda = 0.001$), Set C ($\tau = 100$~fs, $\lambda = 0.01$), Set D ($\tau = 100$~fs, $\lambda = 0.001$), and Set E ($\tau = 20$~fs, $\lambda = 0.001$). We kept the elevated temperature at $T_e=600$~K in this case, for comparison with the benchmarks carried out in Ref.~\citenum{Musil2022}. Therefore, we used 16 beads for PA-$T_e$-CMD at all temperatures.

We evaluated the performance of each spectrum using the Earth Mover’s Distance (EMD) \cite{Rubner1998} as a quantitative metric of spectral similarity. 
\corr{This metric quantifies deviations in the spectra arising from both line positions and lineshapes, the latter being  important, for instance, when decomposing a broad band into contributions from different molecular environments~\cite{Auer2007, Bakker2009}.}
Benchmarking and validation of this metric are provided in ~\corr{Sec}~S10 of the SI, together with the numerical values in Table~S4. By construction, a lower EMD indicates better agreement. For this system, spectra with EMD values within $\sim$80 units are qualitatively indistinguishable, while values exceeding 250 units correspond to clear, visually detectable deviations. Because of the presence of small-amplitude oscillations in the spectra, the threshold for what constitutes a “sufficiently low” EMD can vary somewhat with temperature.

At 600~K (\corr{Fig.}~\ref{fig:benchmarking1}a), all parameter sets reproduce the DVR frequencies and capture the key spectral features, despite their EMD values ranging from 46 (Set~E) to 98 (Set~D). Among them, Set~E (EMD = 47) provides the best agreement, followed by Set~A (EMD = 55).

At 300~K (\corr{Fig.}~\ref{fig:benchmarking1}b), Set~E again performs best (EMD = 35), followed by Set~B (EMD = 83). In contrast, Set~A (EMD = 269) and Set~C (EMD = 275) show more significant deviations across the vibrational features, in line with their much larger EMD scores.

At 50~K (\corr{Fig.}~\ref{fig:benchmarking1}c), all parameter sets avoid the curvature problem, but the spectral lineshapes deteriorate due to a slight overheating of the stretch modes. This heating originates from kinetic energy leakage from the internal ring-polymer modes to the centroid, and becomes particularly severe when $\tau = 100$~fs is used, as it fails to maintain the lower centroid temperature. At this temperature, Set~B (EMD = 552) and Set~E (EMD = 711) provide the most reasonable agreement, while Sets~A, C, and D yield EMD values above 1000, consistent with visibly poorer spectra.

These results establish Set~B and Set~E ($\tau = 10, 20$~fs, $\lambda = 0.001$) as the most robust parameter choices across all temperatures (Table~S5). We therefore employ these sets in the subsequent comparisons with alternative approaches. We note that reducing $T_e$ (e.g., to 350~K) helps mitigate energy leakage into the OH stretch band at low temperatures (\corr{Fig.}~S6), and could permit slightly larger values of $\tau$.

\corr{Fig.}\ref{benchmarking2} compares PA-T$_e$-CMD (Set E)
with T$_e$-PIGS, standard CMD, TRPMD, and classical MD at all three temperatures, using the DVR spectrum as a reference, displayed as gray vertical lines at the position of the vibrational peaks. PA-$T_e$-CMD  accurately captures vibrational peak positions across the temperature range, reproducing the quantum redshift without the overbroadening seen in TRPMD or the spurious curvature-induced redshift of CMD. Its performance closely tracks that of $T_e$-PIGS. A slight thermal broadening and intensity increase of the PA-$T_e$-CMD peaks is observed at  50~K, due to the kinetic-energy leakage mentioned above. This broadening is far smaller than the TRPMD spectrum at the same temperature. \corr{In Fig.~S12 of the SI, we show that  PA-$T_e$-CMD shows accurate lineshapes also for the spectrum of liquid water at room temperature, with the qTIP4P/f model.}

\begin{figure*}[ht!]
\centering
\includegraphics[width=0.9\textwidth]{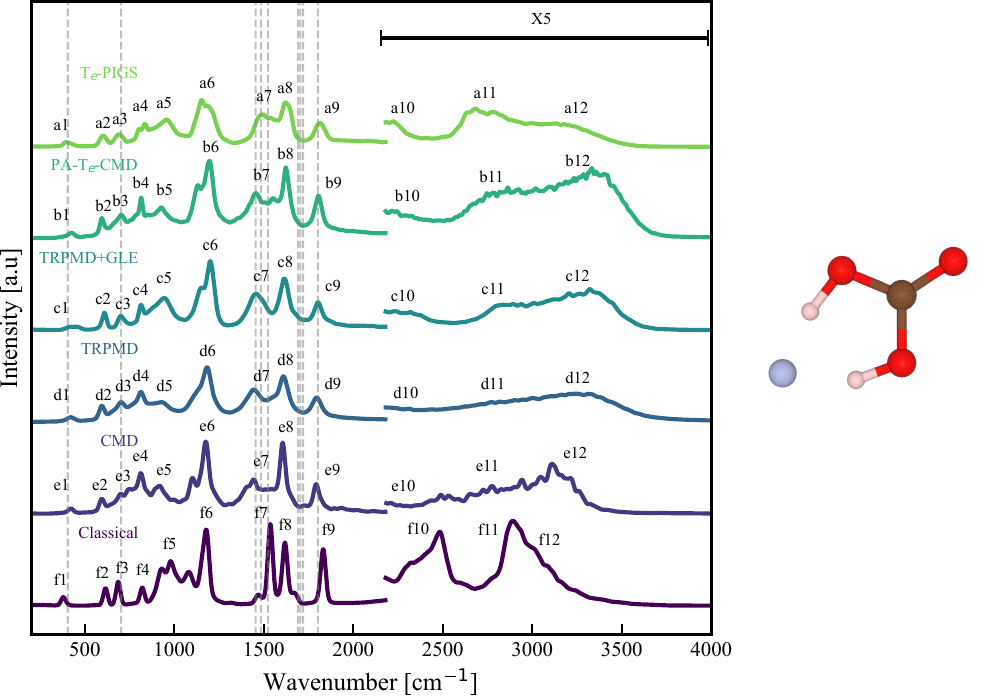}
\caption{ Vibrational density of states (VDOS) of carbonic acid fluoride (CAF) at 100~K, computed using (from top to bottom) $T_e$-PIGS, PA-$T_e$-CMD, TRPMD-GLE, TRPMD, CMD, and classical MD. \corr{$T_e = 350~\text{K}$} for $T_e$-PIGS and PA-$T_e$-CMD.  Experimental IR bands are shown as gray dashed lines.  The molecular structure of CAF is shown on the right. Atom color code: oxygen (red), hydrogen (white), carbon (brown), and fluorine (light blue). 
}
\label{fig:caf_vdos}
\end{figure*}

\subsection{Vibrational Fingerprints of Strongly Anharmonic Vibrations in the Carbonic Acid Fluoride}

Carbonic acid was shown to form a surprisingly stable complex with fluoride atoms in the gas phase, in which strong ionic hydrogen bonds are involved~\cite{Thomas2019}. These bonds in the carbonic acid fluoride [H$_2$CO$_3$F]$^-$ (CAF) molecular complex were shown to exhibit strongly anharmonic dynamics, characterized by fingerprints of its measured and simulated IR spectrum in the OH-stretch region~\cite{Thomas2019}. The most stable conformer is of $C_{2v}$ symmetry, with the fluoride sitting equidistant to both OH groups. Here, we revisit this problem from a simulation perspective and show the performance of different flavours of TRPMD and CMD for the vibrational spectrum.

\corr{Fig.}~\ref{fig:caf_vdos} presents the vibrational density of states (VDOS) for CAF computed at 100~K using $T_e$-PIGS ($T_e$ = \corr{350}~K), PA-$T_e$-CMD ($T_e$ = \corr{350}~K), TRPMD, TRPMD+GLE~\cite{Rossi2018}, CMD and classical MD. It should be noted that the comparison to experiment is qualitative, as the experimental data corresponds to IR active bands at a cryogenic temperature of 0.4~K~\cite{Thomas2019}, which is impractical to simulate with explicit path-integral methods. We prefer to show the full VDOS spectrum for the methodological comparison and note that the calculation of the IR spectrum would not change our conclusions.

In the region below 2000 cm$^{-1}$, all methods yield a fairly similar spectrum. In particular, all methods that include NQEs predict a slight blue-shift of the bands labeled 1 (F$^-$ displacement wrt. carbonic acid) and a slight red-shift of the band labeled 9 (CO stretch coupled to OH bend in-plane bend), with respect to the f1 and f9 bands in the classical-nuclei spectrum. In both cases, this leads to a better agreement with experiment. The TRPMD spectrum presents the largest (spurious) broadening of all peaks, as expected for this method. The bands labeled 6, 7 and 8, comprising C out-of-plane displacement, CO in-plane stretches and OH in and out-of-plane bends are very similar in  PA-$T_e$-CMD, CMD, TRPMD and TRPMD+GLE. $T_e$-PIGS also produces a fairly similar spectrum to the other methods in this region, but with a narrower 7/8 band, which comprises OH in-plane bending motion. The similarity among these methods strongly suggests that spectral broadening is primarily driven by anharmonic coupling.

In the region above 2000~cm$^{-1}$, dominated by symmetric and anti-symmetric OH stretch vibrations, the 
PA-$T_e$-CMD, TRPMD and TRPMD+GLE yield broad spectral features, labeled 10, 11 and 12, with a substantial spectral weight above 3000~cm$^{-1}$, consistent with the anharmonic blue-shift previously reported for this system~\cite{Thomas2019}. 
CMD also exhibits a high-frequency band, but unlike the other methods, it appears red-shifted and slightly more broadened than its PA-$T_e$-CMD counterpart — a signature of the curvature problem. To disentangle the contributions of the curvature problem and kinetic energy leakage to the broadening of the OH stretching bands, we performed a $T_e$-scan using PA-$T_e$-CMD, shown in \corr{Fig.}~S7 (SI). The redshift characteristic of the curvature problem is largely absent already at $T_e = 350$ K; however, the bands remain broad. Since further increasing $T_e$ up to 500 K does not significantly affect the lineshape, this suggests that the observed broadening arises predominantly from the anharmonic dynamics. 
$T_e$-PIGS predicts a different spectral weight of the bands in this region. We will see that this is related to the PMF that was used. 

We trained PMFs for T$_e$-PIGS at 500, 400 and 350~K. We observed that at 500~K an HF molecule forms and promptly dissociates, such that only a few structures of the bound molecule can be obtained from the PIMD simulations. This dissociated structure is a low-energy complex that was discussed in Ref.~\citenum{Thomas2019}\corr{, but we cannot fully exclude that the dissociation could be a result of innacuracies in the underlying MLIP}. Computing a T$_e$-PIGS spectrum with a PMF trained on this data leads to a spectrum that looks very much like the classical one, but with red-shifted high-frequency bands, in disagreement with all other methods (see \corr{Fig.}~S14). Decreasing the temperature allows the molecule to stay bound for a longer time and improves the resulting spectral shape obtained by T$_e$-PIGS. However, the spectral weights on bands 11 and 12 are in slight disagreement with the other methods, likely due to challenges in sampling appropriate centroid configurations for fitting the PMF that are locally similar to those at the lower temperatures. For systems that are weakly bound and reactive, PA-$T_e$-PIGS can thus offer a more robust alternative, as it samples the local PMF on the fly and more faithfully. In addition, PA-$T_e$-PIGS may benefit from some error cancellation, as some energy leakage from the internal modes could increase the spectral weight of the peak labeled 12.

To better understand how anharmonic coupling induces spectral broadening, we analyzed the free energy surfaces (FES) along the O--H and H$\cdots$F$^{-}$ displacement coordinates (\corr{Fig.}~S15). The FES obtained with PA-$T_e$-CMD and TRPMD+GLE are broad, allowing the proton to explore large regions of the potential energy surface at minimal energetic cost (see also \corr{Fig.}~S16). The FES regions explored by the proton in the classical-nuclei simulations are much narrower. The greater mobility in the quantum simulations allows the system to access configurations where the hydrogen--fluorine distances substantially exceed typical hydrogen-bonding lengths, resulting in a loss of the double hydrogen bond present in the optimized structure. Visual inspection confirms that these are structures wherein one \mbox{O--H} group forms a strong single hydrogen bond with fluorine, elongating the bonded O--H, while the other O--H group contracts due to the lack of direct hydrogen-bonding interactions. It is interesting to note that these structures, which break the $C_{2v}$ symmetry, resemble other stable low-energy minima of CAF  discussed in Ref.~\citenum{Thomas2019}. The presence of the free O--H group gives rise to the blue-shifted band in the quantum spectra. In the classical simulations, the F$^-$ anion largely remains equally shared between the two O--H groups.

\subsection{NQEs in the Vibrational Spectra of Different Phases of Methylammonium Lead Iodide (MAPI)}

After assessing the performance of PA-$T_e$-CMD on the water monomer and the CAF molecular complex, we move to the analysis of the vibrational spectrum of the hybrid organic-inorganic perovskite MAPI. The vibrational features of MAPI in its various phases have been studied with the purpose of phase characterization~\cite{Brivio2015, LahnsteinerBokdam2018}, determination of the rotational and reorientation dynamics of organic cations (methylammonium)~\cite{Li2018, Brivio2015}, and understanding of anharmonic lattice vibrations related to its low thermal conductivity~\cite{Gold2018,Tianjun2019,Ashutosh2021}. 
Here, we assess the impact on NQEs on its vibrational spectrum, as predicted by the various flavors of CMD.

\begin{figure*}[ht]
\centering
\includegraphics[width=\textwidth]{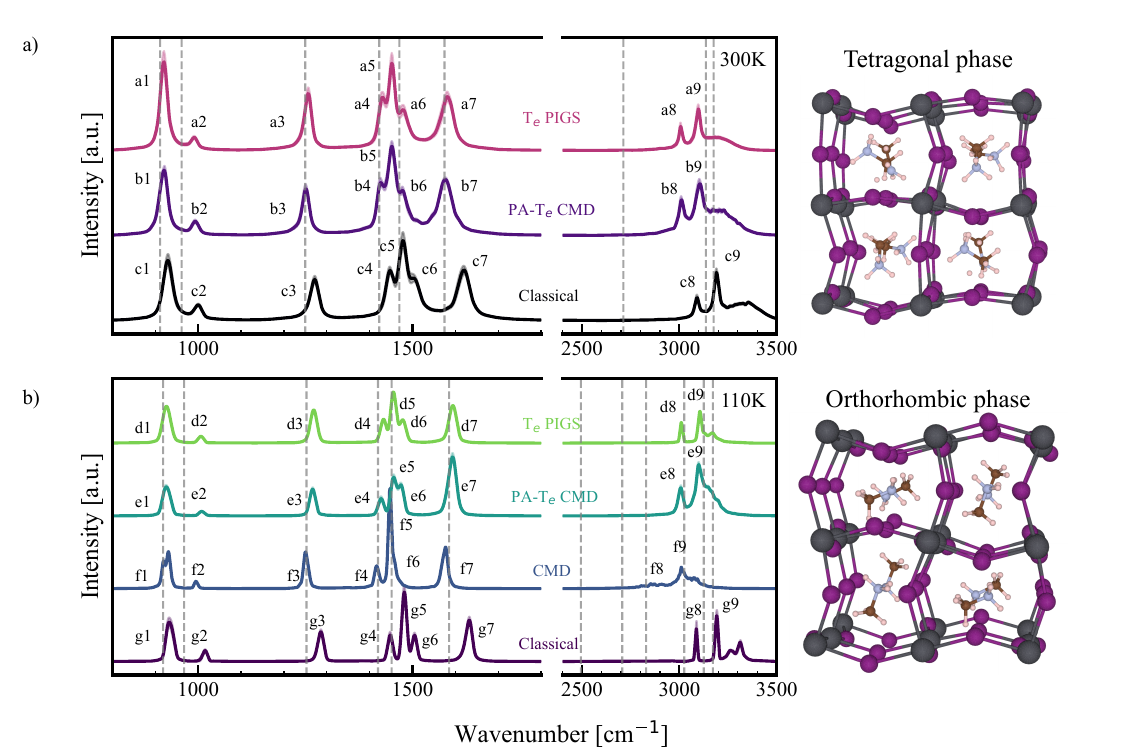}
\caption{
Vibrational density of states (VDOS) of MAPI computed using T$_e$-PIGS, PA-$T_e$-CMD, CMD, and classical MD. 
(a) Tetragonal phase at 300~K. 
(b) Orthorhombic phase at 110~K. 
Experimental IR peak positions are shown as vertical dashed lines for reference. 
The computed spectra are shifted vertically for clarity.
The right panels show the corresponding crystal structures: lead atoms in dark gray, iodine in purple, carbon in brown, nitrogen in light blue, and hydrogen in white. 
}

\label{fig:mapi_vdos}
\end{figure*}

Fig.~\ref{fig:mapi_vdos} shows the vibrational density of states (VDOS) of MAPI in the tetragonal phase at 300 K (a) and the orthorhombic phase at 110 K (b), computed using T$_e$-PIGS ($T_e = 500$~K), PA-$T_e$-CMD ($T_e = 400$~K), standard CMD, and classical MD. The vibrational spectrum of MAPI is typically divided into three regions: (1) low-frequency internal vibrations of the PbI$_3$ network (<100 cm$^{-1}$), (2) methylammonium (MA) cation librations (140–180 cm$^{-1}$), and (3) internal vibrations of the MA cation (800–3100 cm$^{-1}$) \cite{Brivio2015}. Our analysis focuses on the third region, where experimental IR data are available for direct comparison. Experimental IR peak positions from Ref.\citenum{Schuck2018} are shown as vertical gray lines.

At both temperatures, T$_e$-PIGS and PA-$T_e$-CMD exhibit good agreement with the experimental IR spectrum, with most experimental peaks falling within 10~cm$^{-1}$ of the computed ones.
However, deviations of up to 45 cm$^{-1}$ are observed, particularly for features associated with the C–N stretching mode (labeled 2). At 300 K, PA-$T_e$-CMD overestimates this mode by ~30 cm$^{-1}$, similar to T$_e$-PIGS. At 110 K, deviations increase to 44 cm$^{-1}$ for PA-$T_e$-CMD and 38 cm$^{-1}$ for T$_e$-PIGS. Interestingly, classical MD spectra show similar separations between peaks c1/c2 (300 K) and g1/g2 (110 K), suggesting that the discrepancies originate not from the quantum methods, but rather from limitations in the underlying potential, likely related to the level of theory used to train the neural network.

The role of NQEs was assessed by comparing quantum (T$_e$-PIGS, PA-$T_e$-CMD) and classical MD spectra at 300 K. As expected, NQEs are most prominent in the high-frequency region, where the zero-point energy (ZPE) of hydrogen atoms plays a dominant role. For the symmetric N–H stretching mode (labeled 8), redshifts of 83 cm$^{-1}$ (10.4 meV) and 79 cm$^{-1}$ were observed with T$_e$-PIGS and PA-$T_e$-CMD, respectively. The asymmetric N–H stretch (labeled 9) exhibited even larger redshifts of 94 cm$^{-1}$ (11.7 meV) and 85 cm$^{-1}$. Bending modes also showed noticeable redshifts, up to 38 cm$^{-1}$ for T$_e$-PIGS and 44 cm$^{-1}$ for PA-$T_e$-CMD. In contrast, modes dominated by heavier atoms—such as the C–N stretch and CH$_3$NH$_3$ rocking motions (features 1 and 2)—exhibited much smaller shifts ($\sim$9 cm$^{-1}$). Similar trends were observed at 110~K, with quantum effects becoming more pronounced at lower temperatures.

The comparison with CMD is particularly revealing at 110~K, where the curvature problem becomes evident. Features d2 to d9 in T$_e$-PIGS and e2 to e9 in PA-T$_e$-CMD are consistently blue-shifted relative to CMD frequencies.
\corr{In methylammonium, the C--H stretching modes are strongly coupled to rotational motions because the methyl group can freely rotate about the C--N bond. As a consequence, the feature labeled f7 and f8 are significantly red-shifted, with f8 in particular showing extreme broadening. The N--H stretching modes (f9) are also affected by the curvature problem and exhibit redshifts and line broadening, although to a lesser extent than the C--H vibrations. This difference arises because the N--H groups remain hydrogen-bonded to the iodine atoms of the perovskite cage, which restricts full rotation but allows limited librational motion.}
The curvature problem also alters the lineshape, suppressing the asymmetric N--H bending modes (d6 and e6 in the elevated-temperature methods), which merge into a single feature (f5/6) in CMD.

Overall, the T$_e$-PIGS spectra align closely with the PA-T$_e$-CMD spectra, both in position and lineshape. However, in the low-temperature orthorhombic phase, noticeable discrepancies emerge in the high-frequency NH stretching region: the PA-T$_e$-CMD spectra are broader and slightly less resolved compared to their T$_e$-PIGS counterparts. This broadening is attributed to the coupling between physical high-energy modes and the hot non-centroid ring-polymer modes. As shown in \corr{Fig.}~S8, this artifact is mitigated by lowering the elevated temperature even further. However, lowering \( T_e \) below 350~K reintroduces the curvature problem, negatively impacting the spectral accuracy in the low-temperature regime, as expected from our analysis in \corr{Sec.}~\ref{subsec:choosing-te}. 

We note that the orthorhombic phase, in particular, exhibits highly anharmonic energy profiles along the N--H stretch \corr{normal modes}, as discussed in \corr{Sec.}~S13 of the SI. The non-negligible anharmonic coefficients \corr{listed in Table~S6 of the SI} induce stronger coupling between the centroid and internal ring-polymer modes. While PA-$T_e$-CMD partially suppresses this coupling through the adiabatic separation, it is to be expected that spurious effects due to energy exchange between ``hot'' and ``cold'' modes of the ring polymer are more pronounced at the orthorhombic phase.

Finally, we draw attention to the fact that PA-T$_e$-CMD and T$_e$-PIGS largely agree for the tetragonal and orthorhombic  phases of this system, even when the T$_e$-PIGS PMF was parametrized at the high-temperature cubic phase. Therefore, we can conclude that this method is generally successful in capturing NQEs across material phases, presumably as long as the potential energy profile of the high-frequency vibrations is not substantially altered in different phases~\cite{Kapil2024}. 
 As shown in \corr{Sec.}~S14 of the SI, we also investigated the changes on high-frequency O--H stretch peaks of liquid water induced by pressure, as a stringent test of this transferability. While the shifts in peak positions are very small when varying the pressure in the experimentally accessible range, PA-$T_e$-CMD seems to capture the shift in the right direction. $T_e$-PIGS only successfully captures the trend in peak shifts when using elevated temperature PMFs trained at each respective pressure. However, because these changes are extremely small, we refrain from making a strong statement in this respect. In any case, it is plausible that training a general machine-learned PMF for $T_e$-PIGS that is aware of several different material phases will make that method more accurate.

\section{Conclusions}
\label{sec:conclusions}

The findings discussed in this paper reinforce the applicability of the elevated-temperature \textit{ansatz} in CMD as an accurate framework for including NQEs into simulations of vibrational spectra. Compared to the original $T_e$-PIGS implementation, the method developed in this work, the partially-adiabatic PA-$T_e$-CMD, enhances transferability and makes the method simpler to use by eliminating the need for coarse-grained PMFs.
This method emerges as a promising alternative for accurate vibrational calculations suitable for systems with significant configurational flexibility and reactivity, such as molecular complexes, biomolecules, and materials near phase transitions.

The benchmarks we have carried out showed  that the thermostat parameters used in PA-$T_e$-CMD can have a substantial effect on the resulting vibrational spectra. Here, we have explored different parameter regimes of the 2T-PILE thermostat, introduced in \corr{Sec.}~\ref{subsec:implementation}. The $\lambda$ parameter, which controls the degree of damping of Langevin thermostats coupled to the non-centroid ring-polymer modes can be safely chosen in the underdamped regime, with $\lambda = 0.001$ minimizing spurious features in the spectra. The parameters of the thermostat acting on the centroid are also crucial for accurate temperature control in PA-$T_e$-CMD simulations. PA-$T_e$-CMD requires stronger coupling to prevent unwanted energy transfer between the centroid and non-centroid modes. For example, in the water monomer at 50~K, larger $\tau$ allows kinetic energy to leak from ring-polymer internal modes to the high-frequency physical vibrational modes, distorting the vibrational spectrum.  Especially at low temperatures, a smaller thermostat relaxation time $\tau$ of 10--20~fs performs best. \corr{As shown in the benchmarking section, and later in the analysis of the CAF molecule, with this set of parameters, any remaining vibrational energy leakage can only slightly increase the intensity of high-frequency bands.}

The selection of the elevated temperature $T_e$ is equally critical. It must be high enough to mitigate curvature issues while ensuring that high-frequency normal modes, which are most affected by NQEs, remain in their quantum ground state. \corr{In addition, the choice of $T_{e}$  in PA-$T_{e}$-CMD influences the extent of thermal energy leakage and, consequently, the relative intensities of high-frequency modes.} 
We have devised a practical procedure for choosing $T_e$ at different physical temperatures, and confirmed that values of $T_e$ between 350 and 600~K are effective for this method. The optimal value of $T_e$ will depend both on the anharmonic profile of the high-frequency vibrational modes and the physical temperature. \corr{For PA-$T_e$-CMD, using a lower $T_e$ at lower physical temperatures reduces thermal energy transfer from the internal modes to the centroid.}

The two elevated-temperature approaches,  PA-$T_e$-CMD and $T_e$-PIGS, exhibit distinct strengths depending on the system. $T_e$-PIGS is completely adiabatic, but requires a precomputed centroid PMF, which can be computationally demanding to obtain and which can contain fitting errors. Generating a new PMF requires additional data collection and reparameterization, making it more demanding for applications where thermodynamic conditions change. By contrast, PA-$T_e$-CMD eliminates the need for precomputed PMFs and instead computes the centroid PMF on the fly, reducing the setup complexity. However, PA-$T_e$-CMD incurs a higher per-step cost due to the small time-step necessary to propagate the equations of motion, and can suffer from energy leakage. In terms of computational cost, PA-$T_e$-CMD is roughly an order of magnitude more expensive than TRPMD, primarily due to the smaller time step required to maintain adiabatic separation. Its cost is comparable but smaller than standard PA-CMD, as one typically needs  only 8-16 beads to run the simulations at a given $T_e$, for any physical temperature. In contrast, once a suitable machine-learned PMF is available, $T_e$-PIGS offers the fastest dynamics, as it allows for classical time steps and eliminates the need for ring-polymer propagation altogether.

In analyzing vibrational spectra for systems containing multiple atomic species and a varied vibrational landscape, we conclude that both PA-$T_e$-CMD and $T_e$-PIGS efficiently remove the curvature problem of different curvilinear modes in  molecular and condensed-phase systems.
For the CAF molecule, we have clarified the dynamical motions that induce a blue-shifted OH stretching peak when including NQEs. From a methodological perspective, the strongly anharmonic nature of some vibrational modes and the higher reactivity of CAF makes it challenging to train a reliable PMF for $T_e$-PIGS. In this case, PA-$T_e$-CMD indeed presents a more reliable setup. On the other hand, PA-$T_e$-CMD and $T_e$-PIGS yield  very similar vibrational spectra of MAPI tetragonal and orthorhombic phases, with a single PMF trained at the high-temperature cubic phase, and NQEs bring the spectra closer to experimental data. \corr{Despite having presented analyses focused on the VDOS for both MAPI and CAF, we expect our conclusions to hold equally for the IR and Raman spectra of these systems, given the close relationship between VDOS, IR and Raman spectra.}

PA-$T_e$-CMD currently struggles in systems with strong anharmonicity of the potential energy surface and in low-temperature regimes, where the two-temperature thermostat we currently use is unable to maintain the temperature separation between centroid and non-centroid modes. Making PA-$T_e$-CMD spectra more accurate at lower-temperatures and for highly anharmonic systems will require the development of more advanced thermostatting strategies, such as extending the two-temperature thermostating \textit{ansatz} to generalized Langevin equation thermostats tailored for path-integral simulations.

The implementation of PA-$T_e$-CMD is already available in the i-PI code, requiring only a few additional parameters inputs compared to CMD and TRPMD. This streamlined setup makes it more accessible to general users and allows for direct coupling to existing \textit{ab initio} or ML force calculators. Overall, we confirm that different flavors of elevated-temperature approaches mitigate artifacts inherent to CMD and TRPMD, enhance spectral accuracy, and increase efficiency of simulations. While $T_e$-PIGS is preferred when a reliable machine-learned PMF is available, particularly in strongly anharmonic regimes, PA-$T_e$-CMD provides a flexible and user-friendly alternative in scenarios where training such a model is impractical or data are scarce. Our study clarifies the trade-offs between these two implementations and highlights the broader applicability of elevated-temperature path-integral techniques in quantum vibrational spectroscopy. 

\section*{Supplementary Material}
\corr{Supplementary Material is available, containing characterization of the curvature problem in vibrational spectra, workflow for the determination of $T_e$, additional data about the determination of $T_e$, additional data for validation of the implementation, computational details of molecular dynamics simulations and machine-learning models, supporting data for the discussion of the examples shown in the main text, and data regarding liquid water at different pressures.}   

\begin{acknowledgements}
\corr{M.R. acknowledges funding by the European Union (ERC, QUADYMM, 101169761). G.T. gratefully acknowledges support from the Alexander von Humboldt Foundation, J.C. acknowledges funding from the Max Plack Graduate Center for Quantum Materials. V.K. acknowledges support from UCL's startup funds.
We are grateful for computational support from the Swiss National Supercomputing Centre (CSCS) under project s1288 on Alps and UCL Myriad High Performance Computing Facility (Myriad@UCL). This work used computing equipment funded by the Research Capital Investment Fund (RCIF) provided by UKRI, and partially funded by the UCL Cosmoparticle Initiative. We also thank the Max Planck Computing and Data Facility (MPCDF) for enabling the vast majority of the computations presented in this work.} 
\end{acknowledgements}

\section*{Data Availability Statement}

\corr{The data supporting the findings of this study are available in \url{https://doi.org/10.5281/zenodo.17395334}.}

\section*{References}
\bibliographystyle{aipnum4-2}
\bibliography{bibliography}

\end{document}


\title{Supplementary Information for ``Vibrational Spectra of Materials and Molecules from Partially-Adiabatic Elevated-Temperature Centroid Molecular Dynamics"}

\author{Jorge Castro}
\affiliation{MPI for the Structure and Dynamics of Matter, Hamburg, Germany}

\author{George Trenins}
\affiliation{MPI for the Structure and Dynamics of Matter, Hamburg, Germany}

\author{Venkat Kapil}
\affiliation{Department of Physics and Astronomy, University College London, 7-19 Gordon
St, London WC1H 0AH, UK}
\affiliation{Thomas Young Centre and London Centre for Nanotechnology, 9 Gordon St, London WC1H 0AH}

\author{Mariana Rossi}
\email{mariana.rossi@mpsd.mpg.de}
\affiliation{MPI for the Structure and Dynamics of Matter, Hamburg, Germany}

\date{\today}
\maketitle




\section*{Abbreviations}

\begin{enumerate}
    \item[IR] Infrared
    \item[VDOS] Vibrational density of states
    \item[MD] Molecular dynamics
    \item[AIMD] \textit{ab initio} molecular dynamics
    \item[PIMD] Path-integral molecular dynamics
    \item[CMD] Centroid molecular dynamics
    \item[RPMD] Ring polymer molecular dynamics
    \item[CAF] Carbonic acid fluoride
    \item[MAPI] Methylammonium lead iodide
    \item[DVR] Discrete Variable Representation
    \item[DFT] Density functional theory 
    \item[MLIP] Machine learning interatomic potential
    \item[vdW] van der Waals 
    \item[GLE] Generalized Langevin equation
    \item[EMD] Earth mover's distance
\end{enumerate}

\newpage

\section{Curvature problem in water monomer IR spectra}

\begin{figure}[H]
    \centering
    \includegraphics{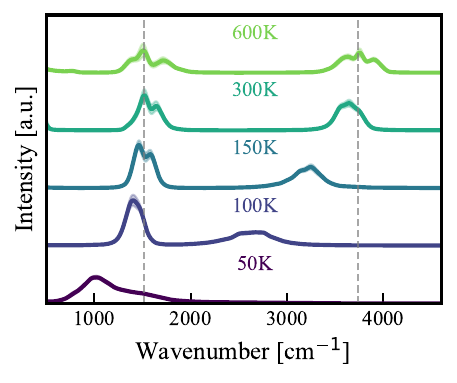} 
    \caption{The IR spectrum of a water monomer described by the Partridge--Schwenke potential energy surface~\cite{PS1997} estimated using CMD at five selected temperatures. The curvature problem manifests as a pronounced red shift in the O--H stretching band upon reduction in temperature and in the bending mode to a lesser extent. 
    The vertical dashed lines indicates the exact ground state transition frequencies for the bending and one of the stretching modes obtained from the numerically exact DVR\cite{Colbert1992} calculations.}
    \label{fig:S1}
\end{figure}

\begin{figure}[H]
    \centering
    \includegraphics{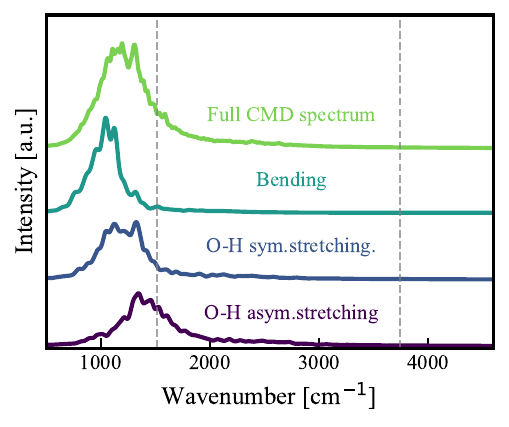} 
    \caption{Projection of the centroid molecular dynamics (CMD) vibrational density of states (VDOS) of a water monomer, described by the Partridge--Schwenke potential energy surface~\cite{PS1997}, at 50~K onto its harmonic normal modes.  The full CMD spectrum (top) is decomposed into contributions from the bending, O--H symmetric-stretching, and O--H asymmetric-stretching modes.  Spectra are vertically shifted for clarity.  Vertical dashed lines indicate the corresponding vibrational frequencies obtained from an exact reference spectrum computed using the discrete variable representation (DVR) method \cite{Colbert1992}.
}
    \label{fig:S2}
\end{figure}

Fig.~\ref{fig:S1} shows a pronounced temperature-dependent red-shift in the IR spectrum of a water monomer estimated using CMD on the Partridge--Schwenke potential energy surface~\cite{PS1997}, resulting in a single band at 50\,K. To demonstrate that this single band at 50\,K arises from the red shift of the stretching mode due to the curvature problem, we projected the IR spectrum obtained from the Partridge--Schwenke potential onto harmonic normal modes and selected those corresponding to the bending and O--H stretching motions (see Fig.~\ref{fig:S2}). Before performing these projections, each instantaneous configuration was aligned to a reference geometry using the Kabsch algorithm \cite{Kabsch:a12999} to remove overall rotational motion. The results confirm that the three IR-active vibrational modes merge into a broad feature in the CMD spectrum at low temperature. The O--H stretching bands exhibit a red shift exceeding 2000~cm$^{-1}$, underscoring the severity of the curvature problem at low temperatures. 
Details of the simulation protocol are provided in Section~\ref{extra_methods} of the Supporting Information.

\newpage

\section{CMD Mean-Field Forces and Critical Radius $r_x$}
\begin{figure}[ht]
    \centering
    \includegraphics[width=0.55\textwidth]{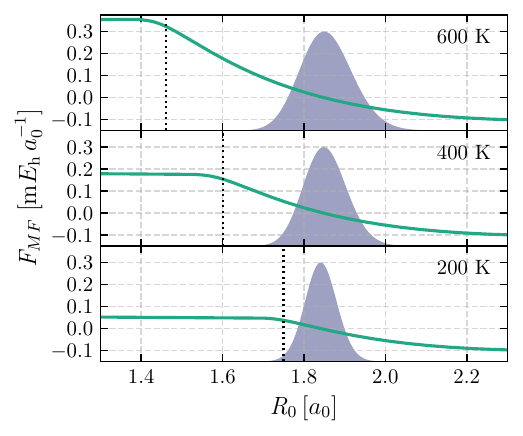}
    \caption{The centroid mean-field force associated with the centroid radial coordinate $R_0 = \sqrt{X^{2}_{0}+Y^{2}_{0}}$ of a two-dimensional `champagne-bottle' model of a vibrating and rotating O--H bond (green line) is shown together with the corresponding CMD Boltzmann distribution of the same coordinate (shaded purple). The dotted black vertical lines mark the critical radius $r_x$, defined in Eq.~(10) of the main text. This data was produced by G. Trenins for the publication in Ref.~\citenum{TreninsJCP2018} and has been replotted here for clarity.
}
    \label{fig:S3}
\end{figure}

\newpage

\section{Workflow for Determining the Elevated Temperature $T_e$ from Bond Scans}
\begin{figure}[ht]
    \centering
    \includegraphics[width=0.95\textwidth]{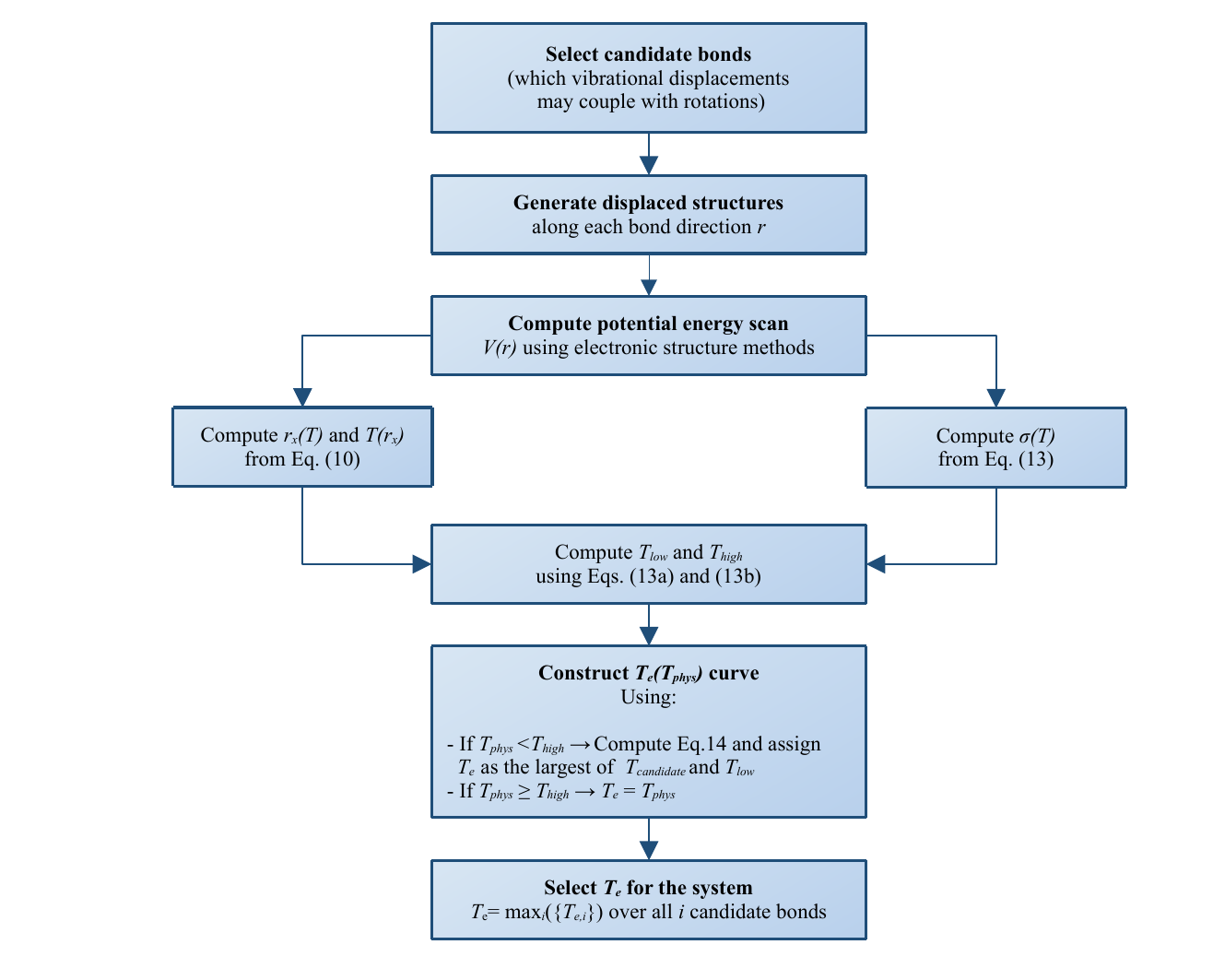}
    \caption{Workflow for determining the elevated temperature $T_e $ using bond-based potential energy scans. Each selected bond is displaced along its bond direction to construct the one-dimensional potential energy surface $V(r)$. From this potential, the crossover radius $r_x(T)$ and distribution width $\sigma(T)$ are respectively evaluated from Eqs.~(10) and~(12) of the main text, while the temperature bounds $T_\mathrm{low}$, $T_\mathrm{high}$, and the candidate temperature $T_\mathrm{candidate}$  are determined from Eqs.~(13a), (13b) and~(14), respectively, of the main text. The optimal $T_e$ is chosen as the highest among all bond-specific results.
    }
    \label{fig:S4}
\end{figure}

\section{Determination of the Elevated Temperature $T_e$ for Vibrational Coordinates in MAPI}

\begin{figure}[H]
    \centering
    \includegraphics[width=0.55\textwidth]{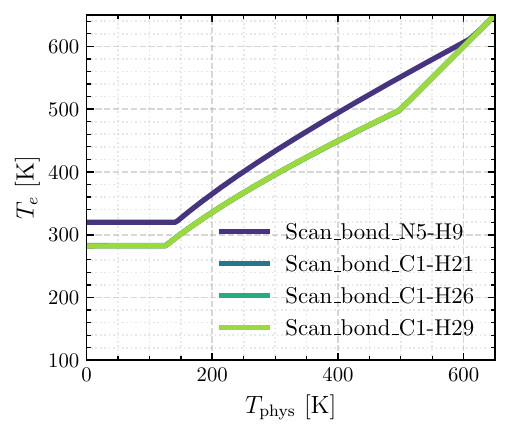}
    \caption{Determination of the elevated temperature $T_e$ for the C--H and N--H bonds of MAPI. The one-dimensional potential energy profiles $V(r)$ for each bond were obtained by displacing the light atom along the bond direction in small increments around the equilibrium geometry and computing single-point energies for each configuration. The elevated temperature $T_e$ was then determined following Eqs.~(10)--(14) of the main text. Among the bonds examined, the N--H bond yields the highest $T_e$}
    
    \label{fig:S5}
\end{figure}

To determine representative elevated temperatures $T_e$ for MAPI, we selected one N–H bond and the three C–H bonds of the methylammonium cation. At 110~K, the N–H groups are stably hydrogen-bonded to the iodine atoms of the perovskite cage. The methyl group undergoes nearly free rotation about the C–N bond. We determined $T_e$ individually for each of them following the procedure described in section II.C of the main text. The selected bonds are labeled as N9–H9, C1–H21, C1–H26, and C1–H29. The plot in Fig.~\ref{fig:S5} shows that there is no considerable distinction among the three C–H bonds. As summarized in Table~\ref{tab:Te_curves}, all three lead to similar values of $T_e$, which are slightly lower than that obtained for the N–H stretching coordinate.

\begin{table}[h!]
\centering

\begin{tabular}{lccccc}
\hline
Label & $r_{\mathrm{eq}}$ (\AA) & $\mu$ (a.u.) & $\omega_{\mathrm{harm}}$ (cm$^{-1}$) & $T_{\min}$ (K) & $T_e(110$ K) \\
\hline
Scan\_bond\_N5-H9 & 1.0200 & 1714.61 & 3259.2 & 320.0 & 320.0 \\
Scan\_bond\_C1-H21 & 1.0850 & 1696.49 & 3138.4 & 282.2 & 282.2 \\
Scan\_bond\_C1-H26 & 1.0850 & 1696.49 & 3154.9 & 282.8 & 282.8 \\
Scan\_bond\_C1-H29 & 1.0850 & 1696.49 & 3138.4 & 282.2 & 282.2 \\
\hline
\end{tabular}
\caption{Summary of $T_e(T_{\mathrm{phys}})$ computations for selected vibrational coordinates in MAPI.}
\label{tab:Te_curves}
\end{table}

\newpage
\section{Elevated temperature tests}
\subsection{Water monomer}
\begin{figure}[H]
    \centering
    \includegraphics[width=0.55\textwidth]{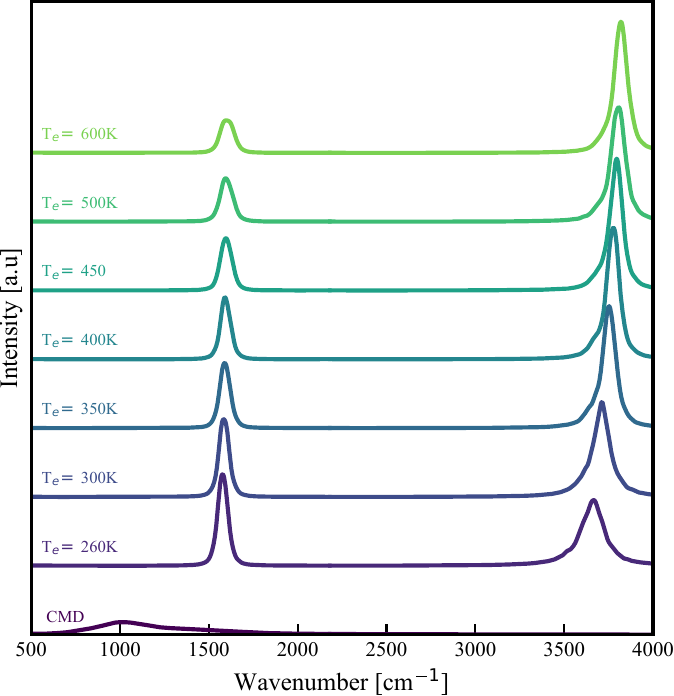}
    \caption{
    IR spectrum of the water monomer described by the Partridge--Schwenke potential energy surface at 100 K computed using CMD and PA-$T_e$-CMD at various elevated temperatures $T_e$, with fixed thermostat parameters $\tau = 100$ fs and $\lambda = 0.01$ (referred to as Set C in the main text). CMD corresponds to the limiting case $T_e = T = 100$ K. As $T_e$ is lowered, the excessive intensity in the O--H stretching band caused by kinetic energy leakage is reduced. However, overly low values of $T_e$ reintroduce the curvature problem, highlighting the trade-off between suppressing leakage and avoiding spurious red shifts.}
    \label{fig:S6}
\end{figure}

\newpage

\subsection{CAF}
\begin{figure}[H]
    \centering
    \includegraphics[width=0.65\textwidth]{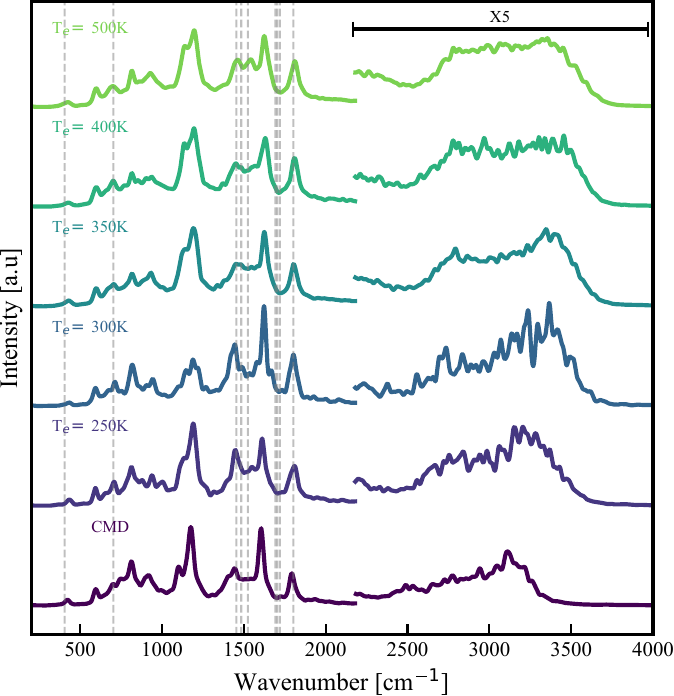} 
   \caption{
    IR of the CAF molecule described using an MLIP (detailed in Sec. \ref{sec:mace-pots}) at 100 K computed using CMD and PA-$T_e$-CMD with increasing elevated temperatures $T_e$ (thermostat parameters fixed to Set C: $\tau = 100$ fs, $\lambda = 0.01$). In standard CMD ($T_e = T = 100$ K), the O--H stretching modes are significantly broadened and red-shifted due to the curvature problem. These artifacts persist up to $T_e = 300$ K, but are largely absent at $T_e = 350$ K and above, indicating suppression of the curvature problem.
    }
    
    \label{fig:S7}
\end{figure}

\newpage

\subsection{MAPI}
\begin{figure}[H]
    \centering
    \includegraphics[width=0.8\textwidth]{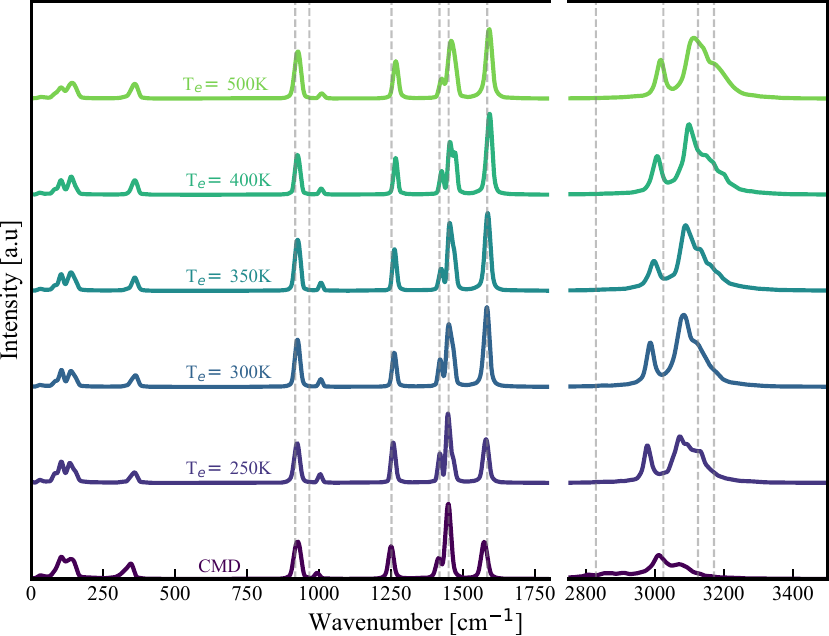}
    \caption{
    Comparison of the VDOS  of orthorhombic MAPI at 110\,K using an MLIP (detailed in Sec. \ref{sec:mace-pots}), computed using CMD and and PA-$T_e$-CMD with increasing elevated temperatures $T_e$. Two spectral regions are shown: the fingerprint region (0–1800\,cm$^{-1}$) and the high-frequency stretching region (2750–3500\,cm$^{-1}$). Vertical dashed lines mark selected experimental IR peak positions. 
    The N--H stretching modes (3100–3300~cm\(^{-1}\)) and C--H stretching modes (2850–3000~cm\(^{-1}\)) are spuriously broadened and red-shifted due to the curvature problem in the CMD spectrum. As $T_e$ increases, spurious red shifts and broadening are suppressed, and at $T_e = 350$\,K, signatures of the curvature problem are largely absent. Further increasing $T_e$ promotes kinetic energy leakage from internal ring-polymer modes to the centroid, which degrades spectral resolution—particularly in the N–H stretching region for $T_e = 500$~K.}
    \label{fig:S8}

\end{figure}

\newpage
\section{Validation of the 2T-PILE Thermostat in Extended Systems}
\begin{figure}[H]
    \centering
    \includegraphics[width=0.55\textwidth]{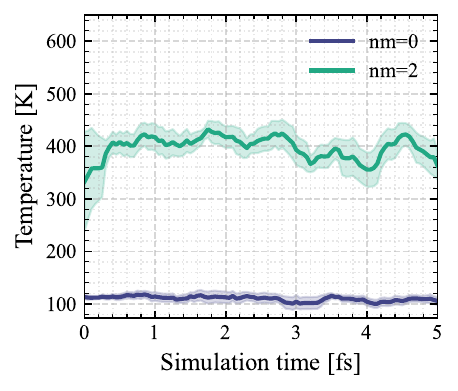}
    \caption{Time evolution of the centroid (nm = 0) and the second internal (nm = 2) normal-mode temperatures during a PA-$T_e$-CMD simulation of MAPI. The centroid mode remains thermostatted near the physical temperature (110~K), while the internal modes thermalize at the elevated temperature (400~K), confirming that the 2T-PILE thermostat maintains independent temperature control in extended systems. Solid lines represent centered moving averages, and shaded areas denote 95\% confidence intervals computed from a moving window of 10 data points (\(\approx 0.5\)~fs)}
    \label{fig:S9}

\end{figure}

\newpage
\section{Computational Details for PA-$T_e$-CMD Benchmarking}

While PA-T$_e$-CMD would be a consistent choice for generating thermalized starting configurations, in this study we employed PIMD instead. As shown below, the results confirm that this choice does not affect the subsequent spectral analyses.
To validate this choice, we computed the IR spectra of a water monomer described by the Partridge–Schwenke potential energy surface at three temperatures using PA-T$_e$-CMD for production, with initial configurations obtained from (i) PIMD with a standard PILE-L thermostat ($\tau=10$ fs, $\lambda=0.01$, time step = 0.25 fs) and (ii) PA-T$_e$-CMD thermalization using a two-temperature PILE-L thermostat with a time step of 0.01~fs ($\tau=10$ fs, $\lambda=0.01$, $\omega=24000$ cm$^{-1}$). 

\begin{figure} []
    \centering
    \includegraphics[width=0.9\textwidth]{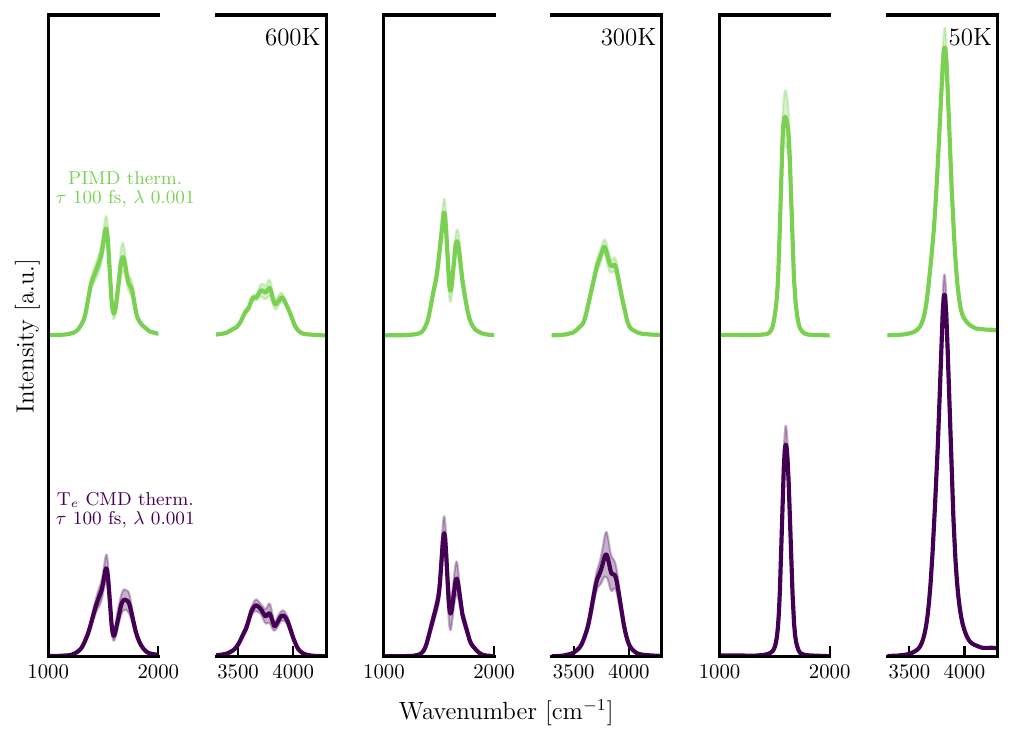} 
    \caption{IR spectra of water monomer described by the Partridge–Schwenke potential energy surface computed using PA-T$_e$-CMD for production runs, from different thermalization methods: PIMD and PA-T$_e$-CMD with different thermostats.}
    \label{fig:S10}
\end{figure}

As shown in Fig.~\ref{fig:S10}, the IR spectra obtained from these two thermalization procedures are nearly identical, indicating that the choice of thermalization method does not affect the subsequent spectral analysis. 

Although PA-$T_e$-CMD thermalization can, in principle, accelerate equilibration—especially at low temperatures where fewer beads are required due to the elevated internal-mode temperature—it also demands a substantially smaller integration time step. For the temperature range examined here (50–600~K), this trade-off results in PIMD thermalization being faster in wall-clock time for the systems considered, and thus it was adopted for the more complex systems discussed in the main paper.

Unless stated otherwise, spectral analyses used a lag time of 5000~fs, 15000 zero-padding steps, a Hanning window \cite{Harris1978}, and a sampling stride of 1 fs.

\subsection{Water monomer} \label{subsec:Comp_details_Bench}

A water monomer described by the Partridge–Schwenke potential~\cite{PS1997} served as a benchmark for PA-T$_e$-CMD and $T_e$-PIGS at 600~K, 300~K, and 50~K. Simulations consisted of 100 ps thermalization followed by 100 independent 10 ps production runs. For PA-\( T_e \)-CMD, we used 16 beads and \( T_e = 600 \)~K, consistent with prior studies~\cite{Musil2022, Kapil2024, Zaporozhets2024}. Thermalization was performed using a PILE-L thermostat (\( \tau = 10 \)~ps) and a 0.5~fs time step.  A 0.01~fs time step was used in production, and data were recorded every 0.5~fs. IR spectra were obtained via Fourier transforms of the dipole–moment time derivative autocorrelation function, using a lag time of 1500~fs, 3000 zero-padding steps and sampling stride of 1 fs.

\subsection{MAPI}

Calculations of MAPI were performed at 110 K and 300 K using a $3 \times 3 \times 3$ supercell, with initial structures optimized to match neutron powder diffraction data \cite{Weller2015}. Simulations employed a MACE potential \cite{Batatia2022mace, Batatia2022Design} trained on HSE06+vdW DFT data \cite{Heyd2003, Krukau2006, Tkatchenko2012} (see Sec. \ref{sec:mace-pots} for details).
PIMD thermalization lasted 100 ps with a PILE-L thermostat ($\tau = 10$~fs) and a 0.25 fs time step. From these, 10 configurations were selected for PA-T$_e$-CMD. Production simulations ran for 10 ps using a 0.025 fs time step, $\omega = 24000$~cm$^{-1}$, $\tau = 20$~fs, $\lambda = 0.001$, and $T_e = 500$~K with 8 beads.

\subsection{CAF}

We modeled CAF using a MACE potential trained on B3LYP+vdW \cite{Becke1993}\cite{Stephens1994}\cite{Tkatchenko2009} DFT data (see Sec. \ref{sec:mace-pots} for details). 
PIMD thermalization was performed at 100 K for 200 ps with 64 beads, a PILE-L thermostat ($\tau = 10$~fs), and a 0.25 fs time step. From these, 50 configurations were selected for PA-T$_e$-CMD. Production runs lasted 10 ps using a 0.025 fs time step, with $\omega = 24000$~cm$^{-1}$, $\tau = 20$~fs, $\lambda = 0.001$, $T_e = 500$~K, and 8 beads.

\subsection{Liquid Water at Different Pressures}

We computed the VDOS of liquid water at 300 K and pressures of 0.05 to 0.40 GPa using the qTIP4P/f potential~\cite{Habershon2009}. Initial configurations of 500 molecules were generated using PACKMOL \cite{Martinez2009}. For each pressure, 2 ns PIMD equilibration in the NPT ensemble was conducted with 32 beads, an isotropic barostat \cite{Bussi2009} ($\tau = 100$~fs), a Langevin thermostat ($\tau = 100$~fs), and a PILE-L thermostat ($\tau = 10$~fs). The average lattice parameters and densities obtained from classical MD and PIMD equilibrations are reported in Tables~\ref{tab:liquid_water_MD} and~\ref{tab:liquid_water_PIMD}, respectively. Production simulations ran for 200~ps in NVT, followed by 200 independent PA-T$_e$-CMD runs of 10 ps each, with $\omega = 24000$~cm$^{-1}$, $\tau = 10$~fs, $\lambda = 0.001$, $T_e = 600$~K, and 32 beads, using the q-TIP4P/F \cite{Habershon2009} water model.

\begin{table}[h]
    
    \centering
    \begin{tabular}{|c|c|c|}
        \hline
        Pressure [GPa] & Cell length [\AA] & $\rho$ [g/cm$^3$] \\
        \hline
        0.05 & 24.49268 & 1.01697 \\
        \hline
        0.40 & 23.66379 & 1.12594 \\
        \hline
    \end{tabular}
    \caption{Equilibrated simulation cell dimensions and corresponding densities for liquid water at various pressures, obtained from 2\,ns equilibration NPT classical MD simulations.}
    \label{tab:liquid_water_MD}
\end{table}

\begin{table}[h]
    
    \centering
    \begin{tabular}{|c|c|c|}
        \hline
        Pressure [GPa] & Cell length [\AA] & $\rho$ [g/cm$^3$] \\
        \hline
        0.05 & 24.48006 & 1.01725 \\
        \hline
        0.40 & 23.68406 & 1.12469 \\
        \hline
    \end{tabular}
    \caption{Equilibrated simulation cell dimensions and corresponding densities for liquid water at various pressures, obtained from 2\,ns equilibration NPT PIMD simulations.}
    \label{tab:liquid_water_PIMD}
\end{table}

\newpage

\section{Training of MACE Potentials \label{sec:mace-pots}}

The MLIPs for CAF and MAPI were trained using the MACE framework~\cite{Batatia2022mace,Batatia2022Design}. 
For both systems, total energies and atomic forces were obtained from DFT calculations using the FHI-aims code~\cite{AIMS2024}. 

\subsection{CAF}

The MACE potential for CAF was trained on a dataset of 4000 configurations obtained from \emph{ab initio} simulations performed with i-PI \cite{Litman2024} using FHI-aims as force provider at the B3LYP+vdW functional level of theory \cite{Tkatchenko2012}. 
Specifically, 1000 configurations were sampled from \emph{ab initio} MD trajectories at 100~K, 200~K, and 600~K (each with a 1~fs time step, SVR thermostat, and relaxation time of 500~fs), and 1000 configurations were sampled from \emph{ab initio} PIMD simulations at 300~K using 16 beads, a PILE-G thermostat ($\lambda = 0.5$, $\tau = 500$~fs), and a 1~fs time step. 
Training employed two interaction layers, 64 channels, a maximum angular momentum $L_\mathrm{max}=0$, correlation order 3, and a cutoff radius of 6.0~\AA. 
Energies and forces were weighted by factors of 10 and 1000, respectively, during optimization. 
The dataset was randomly split into 80\% for training and 20\% for testing.  Training used the Adam optimizer \cite{kingma2017adammethodstochasticoptimization} for up to 300 epochs with early stopping.
The final model achieved mean absolute errors of 0.1~meV/atom for energies and 3.4~meV/\AA{} for forces on the test set.

\subsection{MAPI}

The MACE potential for MAPI was trained on DFT data using the HSE06 hybrid functional~\cite{Heyd2003,Krukau2006} and many-body dispersion corrections~\cite{Tkatchenko2012}. 
The training dataset comprised 648 configurations of $3\times3\times3$ supercells containing 48 atoms each. 
These configurations sampled three distinct crystalline phases of MAPI: cubic (216 structures), orthorhombic (223 structures), and tetragonal (210 structures), extracted from \emph{ab initio} molecular dynamics trajectories at representative temperatures.
The neural network architecture was identical to that used for CAF, with two interaction layers, 64 channels, maximum angular momentum $L_\mathrm{max}=0$, correlation order 3, and a 6.0~\AA{} cutoff radius. 
Energies and forces were included in the loss function with respective weights of 10 and 1000. 
The dataset was split into training, validation, and test subsets in an 80:10:10 ratio, and the Adam optimizer was employed.
The final model achieved mean absolute errors of approximately 0.3~meV/atom for energies and 13.6~meV/\AA{} for forces on the test set.

\newpage

\section{Computational Details for Additional Molecular Simulation Methods} \label{extra_methods}

All simulations, except for the AIMD, used the same interatomic potentials as those employed in the PA-T$_e$-CMD simulations for the corresponding systems. For the water monomer, the Patridge-Schwenke water model \cite{PS1997} was used, while liquid water simulations utilized the q-TIP4P/F model \cite{Habershon2009}. For MAPI and CAF, the MACE potentials \cite{Batatia2022mace} described in Section~\ref{sec:mace-pots} were employed. 
All simulations were performed using the i-PI code\cite{Litman2024}, with FHI-aims~\cite{AIMS2024} serving as the force provider for the AIMD simulations and the corresponding model or machine-learning potentials acting as external drivers in the remaining cases.

\subsection{MD}
Classical MD simulations were conducted by setting the number of replicas to one in i-PI. Simulations began with a 100 ps equilibration run in the NVT ensemble, using an SVR thermostat \cite{Bussi2007,Bussi2008} with a time constant of 10 fs. This was followed by a 100 ps sampling run in the same ensemble and with identical thermostat parameters. From the sampling trajectories, configurations were selected at 1 ps intervals to serve as starting points for independent production simulations. Each production run lasted 10 ps and used an SVR thermostat with a relaxation time ($\tau$) of 500 fs. For MAPI, 10 independent production runs were performed, while 50 were used for CAF, and 200 for liquid water at different pressures. The IR and vibrational density of states (VDOS) spectra were computed using the same parameters as specified in the main text for PA-T$_e$-CMD and T$_e$-PIGS simulations.

\subsection{CMD}
Partially adiabatic CMD simulations were performed for the water monomer at all temperatures studied, for CAF at 100 K and for MAPI at 110 K. These simulations employed a PILE-G thermostat with a time constant of 20 fs and an adiabatic separation parameter of $\Omega = 24000$ cm$^{-1}$. A timestep of 0.025~fs was used, and centroid positions were recorded every 1 fs. 100 independent simulations for the water monomer were run for 10 ps each, with the number of replicas varying depending on the temperature: 16, 32, 64, 128, and 256 replicas for temperatures of 600 K, 300 K, 200 K, 100 K, and 50 K, respectively. For CAF, we run 50 simulations of 10 ps each at 100 K using 64 beads. For MAPI, 10 independent simulations were conducted for 10 ps each using 64 replicas at 110 K.

\subsection{TRPMD}
TRPMD simulations were performed for the water monomer, liquid water, and CAF in the canonical ensemble. A PILE-G thermostat with a time constant of 500 fs and $\lambda = 0.5$ was used. The simulations employed a timestep of 0.25 fs, with bead positions recorded every 0.5 fs. For the water monomer we run 100 independent production runs, as well as for liquid water at different pressures. For CAF we employed 50 independent simulations. Each run lasted 10 ps for all the systems. For the water monomer, simulations were performed at 50 K, 300 K, and 600 K using 256, 32, and 16 replicas, respectively. For liquid water, simulations at 300 K were conducted with 32 replicas, while for CAF, simulations at 100 K used 64 replicas. Additionally, for CAF we employed a GLE thermostat \cite{Rossi2018}, using the GLE(C) parameterization. The corresponding $\boldsymbol{A}$ matrix was obtained from the website~\citenum{gle4md}. 


\subsection{AIMD}
AIMD simulations were conducted solely for CAF at 100 K. The i-PI code was used as the MD engine, while FHI-AIMS was used for force calculations based on DFT at the B3LYP+vdW functional level of theory \cite{Tkatchenko2012}. Two independent equilibration runs were conducted in the NVT ensemble using an SVR thermostat with $\tau = 10$ fs. The final configurations from each equilibration run were used as starting points for two production simulations, each lasting 25 ps. A timestep of 1 fs was employed, with configurations recorded every 1 fs.

\newpage

\section{Spectral Similarity Metric: Earth Mover’s Distance}

To quantitatively evaluate the agreement between simulated and reference spectra, we employed the EMD \cite{Rubner1998}. EMD offers a physically intuitive measure of spectral dissimilarity, as it quantifies the minimum amount of ``work'' needed to transform one distribution into another—interpreting spectra as probability distributions of spectral weight. To validate the robustness of this metric in our context, we first benchmarked EMD using a controlled set of artificial spectra with known displacements and broadenings. 

\begin{figure}[H]
    \centering
    \includegraphics{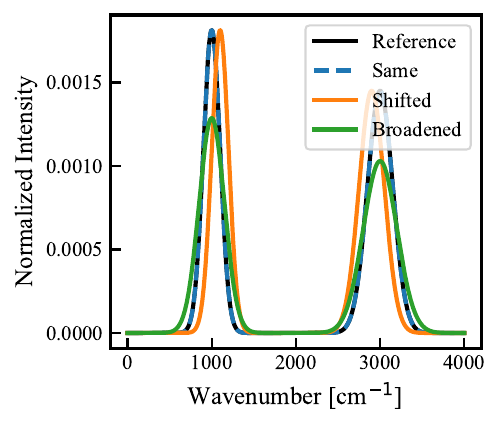} 
    \caption{Benchmark spectra used to validate the EMD as a measure of spectral similarity. The reference spectrum (black) is compared to three test cases: an identical spectrum (blue), a version with shifted peaks (orange), and one with broadened features (green).}

    \label{fig:S11}
\end{figure}

\begin{table}[h]
    \centering
    \begin{tabular}{|c|c|}
        \hline
        Spectrum & EMD index  \\
        \hline
        Identical spectrum & 0.00  \\
        \hline
        Shifted peaks & 100.00\\
        \hline
        Broadened peaks & 81.26  \\

        \hline
    \end{tabular}
    \caption{EMD values computed for the benchmark spectra shown in Fig.~\ref{fig:S11}. As expected, the identical spectrum yields zero distance, while peak shifts and broadenings increase the EMD in proportion to the spectral deformation.}

\end{table}

The results of the EMD analysis applied to all methods and spectral regions discussed in the main text for the water monomer are summarized in Table \ref{tab:emd_water_monomer_sorted}, where lower EMD values indicate closer agreement with the reference spectrum.

\begin{table}[H]
    \centering

\begin{tabular}{|c|c|c|c|c|}
\hline
\textbf{Parameter Set ($\tau\,[\mathrm{fs}]$/$\lambda$)} & \textbf{50 K} & \textbf{300 K} & \textbf{600 K} & \textbf{Total EMD Index} \\
\hline
10/0.001 (Set B)  & 551.98 & 83.19  & 90.28  & \textbf{725.45} \\
20/0.001 (Set E) & 710.87 & 35.40  & 46.74  & 793.01 \\
10/0.01 (Set A) & 1128.06 & 269.43 & 55.07  & 1452.56 \\
100/0.001 (Set D) & 1329.36 & 216.29 & 97.64  & 1643.29 \\
100/0.01 (Set C) & 1830.37 & 275.07 & 93.22  & 2198.66 \\

\hline

\end{tabular}
    \caption{EMD  between simulated and reference spectra for the water monomer at 50 K, 300 K, and 600 K using different values of $\tau$ and $\lambda$ in PA-$T_e$-CMD. Lower values indicate better agreement.}
    \label{tab:emd_water_monomer_sorted}
\end{table}

\newpage

\section{Validation of PA-$T_e$-CMD Line Shapes for Liquid Water}

\begin{figure}[H]
    \centering
    \includegraphics[width=0.48\textwidth]{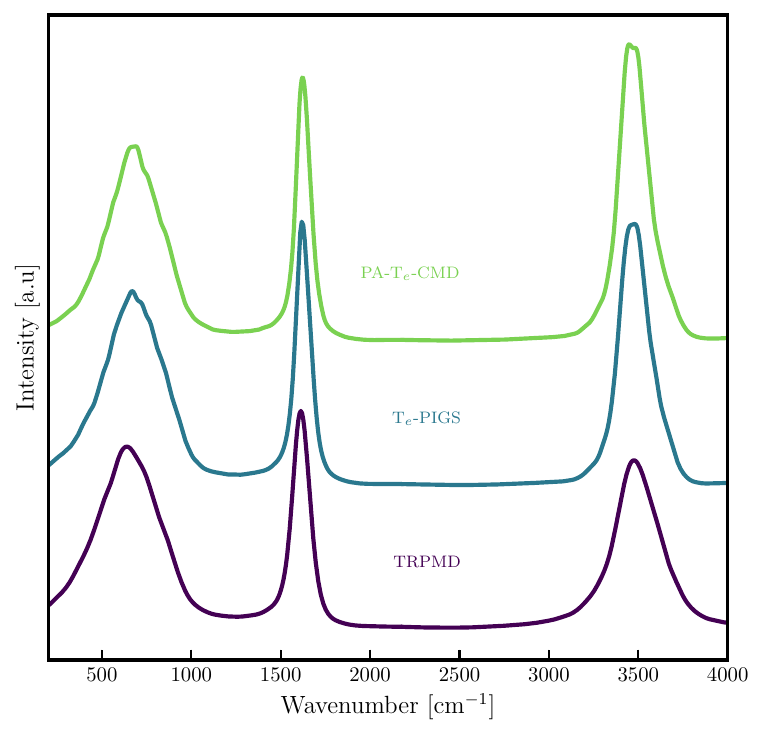}
    \caption{IR of liquid water at 300~K obtained using PA-$T_e$-CMD ($T_e$ = 500~K, $\tau$ = 10~ps, $\lambda$ = 0.001), fully adiabatic $T_e$-PIGS, and TRPMD. The latter two spectra were taken from Ref.~\citenum{Musil2022}. The PA-$T_e$-CMD spectrum closely reproduces the line shapes of the adiabatic $T_e$-PIGS reference while reducing the broadening characteristic of TRPMD. All simulations employ the q-TIP4P/f  water model.\cite{Habershon2009}}
    \label{fig:S12}
\end{figure}

\section{Additional Results: CAF}

\subsection{Assessment of the MACE potential for CAF}

To evaluate the fidelity of the machine-learned MACE potential for CAF, we compared the VDOS spectrum obtained from classical MD simulations at 100~K using the MACE model with that from ab initio MD simulations at the same temperature. As shown in Fig.~\ref{fig:S8}, the spectra are nearly indistinguishable, confirming that the potential accurately reproduces the structural and dynamical features encoded in the reference data. This validates the use of the MACE model for all subsequent quantum simulations of CAF presented in this work.

\begin{figure}[H]
    \centering
    \includegraphics[width=0.48\textwidth]{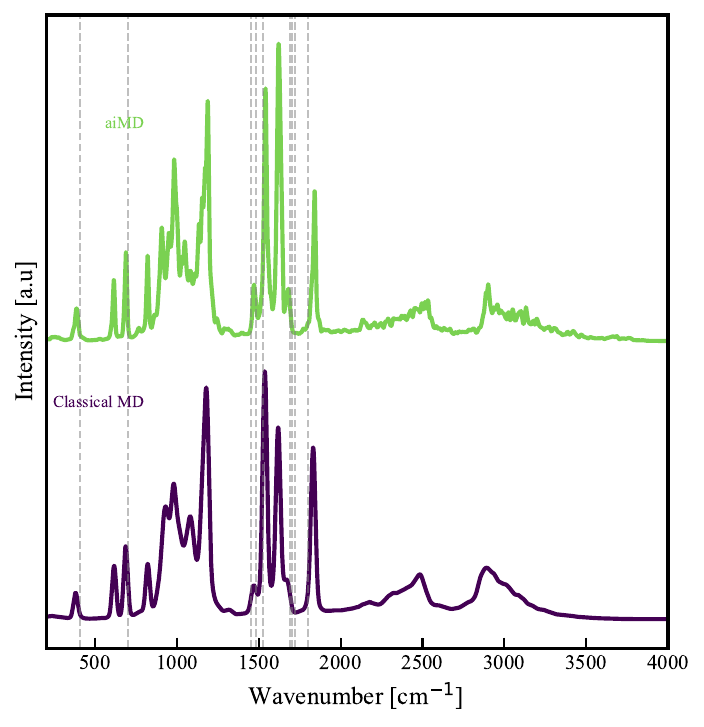}
    \caption{Comparison between the VDOS of CAF computed from AIMD simulations at 100~K and MD simulations using a MACE potential trained on the same AIMD data.}
    \label{fig:S13}
\end{figure}

\subsection{Effect of Elevated Temperatures on $T_e$-PIGS Infrared Spectra of CAF}
\begin{figure}[H]
    \centering
    \includegraphics[width=0.65\textwidth]{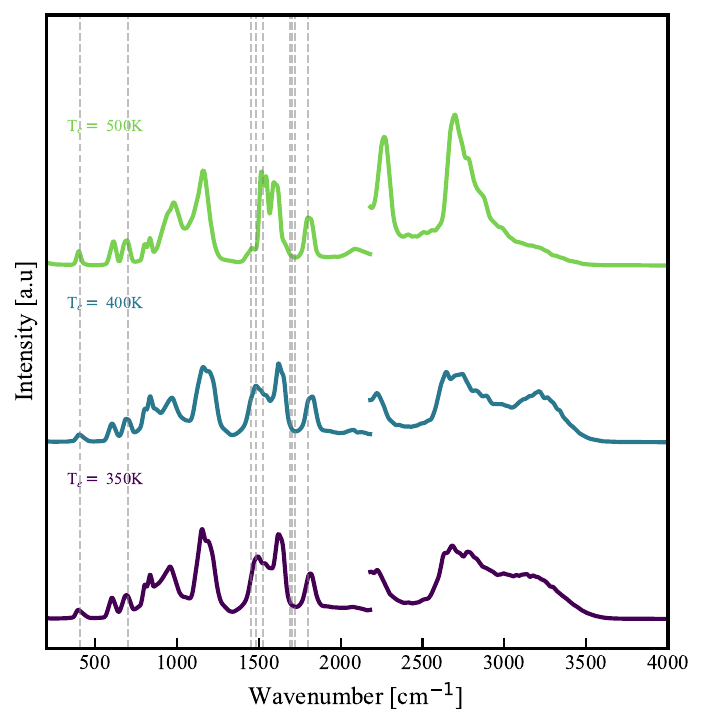} 
   \caption{
    IR spectra of CAF at 100 K computed using $T_e$-PIGS with increasing elevated temperatures $T_e$}
    
    \label{fig:S14}
\end{figure}
\subsection{Free energy surfaces of CAF at 100K}

To compute the free energy surfaces shown below, we extracted O–H and F–H bond distances from trajectory snapshots using MDAnalysis \cite{MDanalysis2011,MDAnalysis2016}. For each hydrogen atom in a given frame, the nearest oxygen atom within a cutoff of 2.0 $\text{\AA}$ was identified and the corresponding O–H distance was recorded. The F–H distance was then computed as the distance between that same hydrogen and the only fluorine atom in the system. 


\begin{figure}[H]
    \centering
    \includegraphics[width=0.9\textwidth]{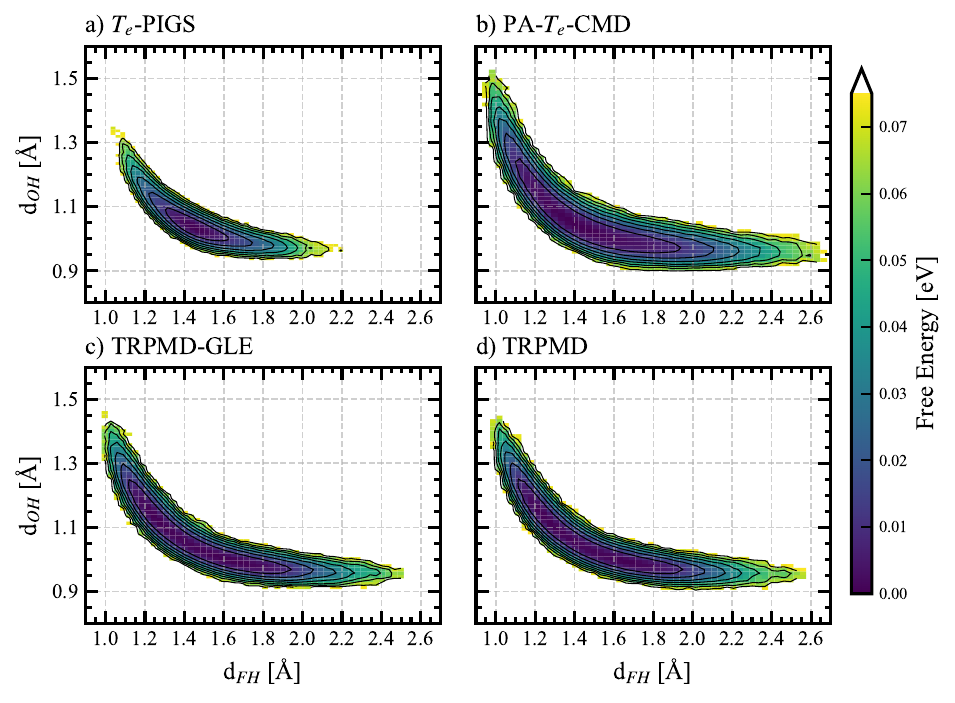} 
    \caption{Free energy surfaces derived from the joint probability density of the O--H and F--H distances. The free energy is calculated as $F = -k_B T \ln \left[P(d_{FH},d_{OH})\right]$, where \(P(d_{FH},d_{OH})\) is the joint probability distribution of the F--H (\(d_{FH}\)) and O--H (\(d_{OH}\)) distances obtained from various PIMD simulation methods at \(T = 100\,\text{K}\). $T_e$-PIGS were simulated with the PMF trained at 500~K. 
    }
    \label{fig:S15}
\end{figure}

\subsection{O-H bond length distribution of CAF with different methods}
\begin{figure}[H]
    \centering \includegraphics{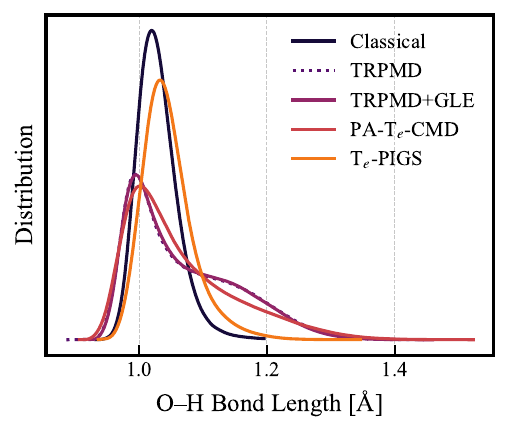} 
    \caption{Normalized O–H bond distributions for carbonic acid fluoride from five simulation methods.} 
    \label{fig:S16}
\end{figure}

\section{Additional Results: MAPI}
\subsection{Comparison of Anharmonic Potentials in Methylammonium Lead Iodide\\
(Orthorhombic Phase at 110\,K vs Tetragonal Phase at 300\,K)}

We fitted the potential energy profiles of the two highest-frequency N–H stretching modes in MAPI, obtained from PA-T$_e$-CMD simulations, using a 5th-order polynomial model of the form
\[
V(x) = a\,x^2 + b\,x^3 + c\,x^4 + d\,x^5,
\]
where \(x\) is the displacement in \AA, \(V(x)\) is the potential energy in eV, and the coefficients \(a\), \(b\), \(c\), and \(d\) have units chosen consistently (See table \ref{tab:nh_stretch_fits}).

\begin{table}[h]
\centering

\begin{tabular}{|l|l|r|r|r|r|}
\hline
Phase & Mode & $a$ (eV/\AA$^2$) & $b$ (eV/\AA$^3$) & $c$ (eV/\AA$^4$) & $d$ (eV/\AA$^5$) \\
\hline
Orthorhombic (110 K) & Symmetric   & 21.29692 & -62.73623 &  28.39502 & 165.88539 \\
Orthorhombic (110 K) & Asymmetric  & 22.74816 &  -1.57264 &  41.84458 &  -0.28620 \\
\hline
Tetragonal (300 K)   & Symmetric   & 12.24795 & -14.36068 &   9.91304 &  -4.83641 \\
Tetragonal (300 K)   & Asymmetric  & 15.44889 &  -0.99118 &  12.89128 &  -1.41118 \\
\hline
\end{tabular}

\caption{Polynomial fitting coefficients for the potential energy profiles of the symmetric and asymmetric N–H stretching modes in orthorhombic (110 K) and tetragonal (300 K) MAPI, obtained from PA-T$_e$-CMD simulations.}
\label{tab:nh_stretch_fits}
\end{table}

\begin{figure}[H]
    \centering
    \includegraphics [width=0.9\textwidth]{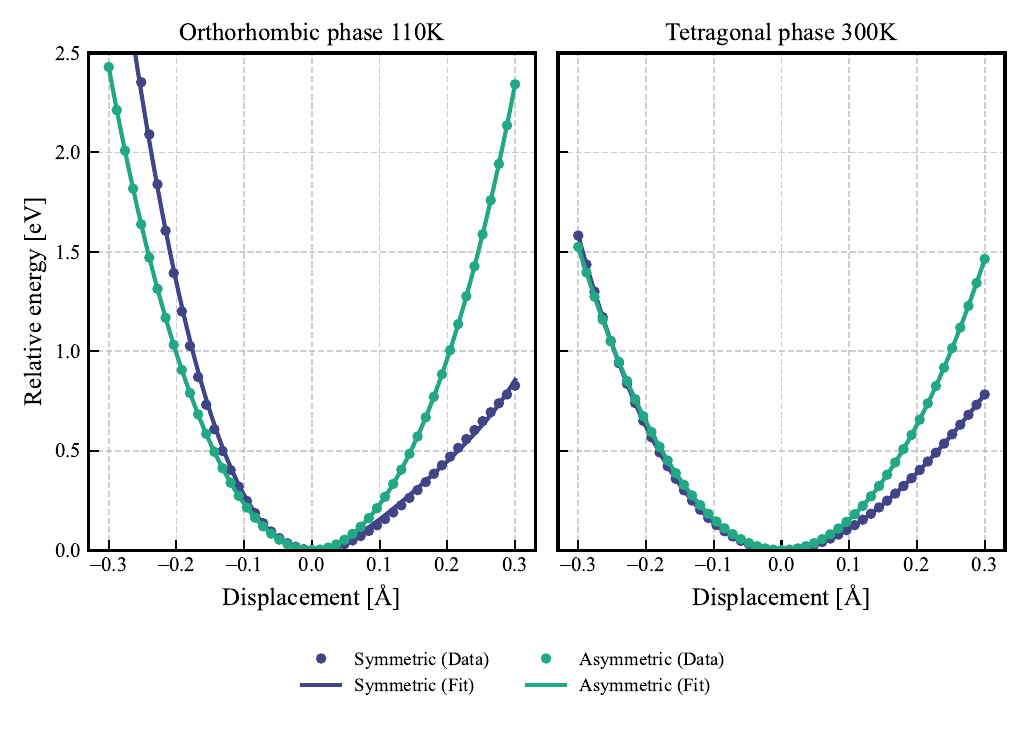} 
    \caption{Extended anharmonic potential fits for N–H stretching in MAPI. Left: Orthorhombic phase at 110K; Right: Tetragonal phase at 300K. Data points (symbols) and polynomial fits (lines) are plotted versus displacement (\AA), with positive displacements indicating bond elongation.}
    \label{fig:S17}
\end{figure}
\newpage

\section{Path-Integral Simulations of High-Pressure Liquid Water: Resolving Subtle Vibrational Shifts}

As a case study, we examine liquid water under high-pressure conditions. Experimental Raman data indicate that increasing pressure from 0.05 GPa to 0.40 GPa strengthens hydrogen bonds, leading to a slight red shift (less than 15 cm$^{-1}$) in the O--H stretching frequency \cite{Okada2005}. This shift arises from the elongation and weakening of the intramolecular O--H bond due to enhanced intermolecular interactions. However, accurately capturing such subtle shifts poses a significant challenge for any computational method.

To investigate these effects, we analyze the vibrational density of states (VDOS) instead of directly computing Raman spectra. In order to accurately deconvolute overlapping features in our simulated VDOS spectra, we fitted the O--H vibrational bands using two skewed Lorentzian functions. Given that the separation between peaks is less than 5\,cm$^{-1}$ and that the individual Lorentzian components are substantially overlapped, a conventional unweighted fit is prone to inaccuracies and potential peak misassignments. To address this challenge, we implemented a dynamic weighting scheme that preferentially emphasizes data points in the peak regions while down-weighting the tails. This approach has been successfully applied to deconvolution of overlapping resonances in NMR spectroscopy \cite{Gipson2006} and determination of spectroscopic parameters from experimental IR absorption spectra \cite{Benner1995, Haaland1985, Johns1987, Chang1977} and is now widely accepted as a standard data reduction methodology \cite{Bevington2003}.

\begin{figure*}[ht]
\centering
\includegraphics[width=\textwidth]{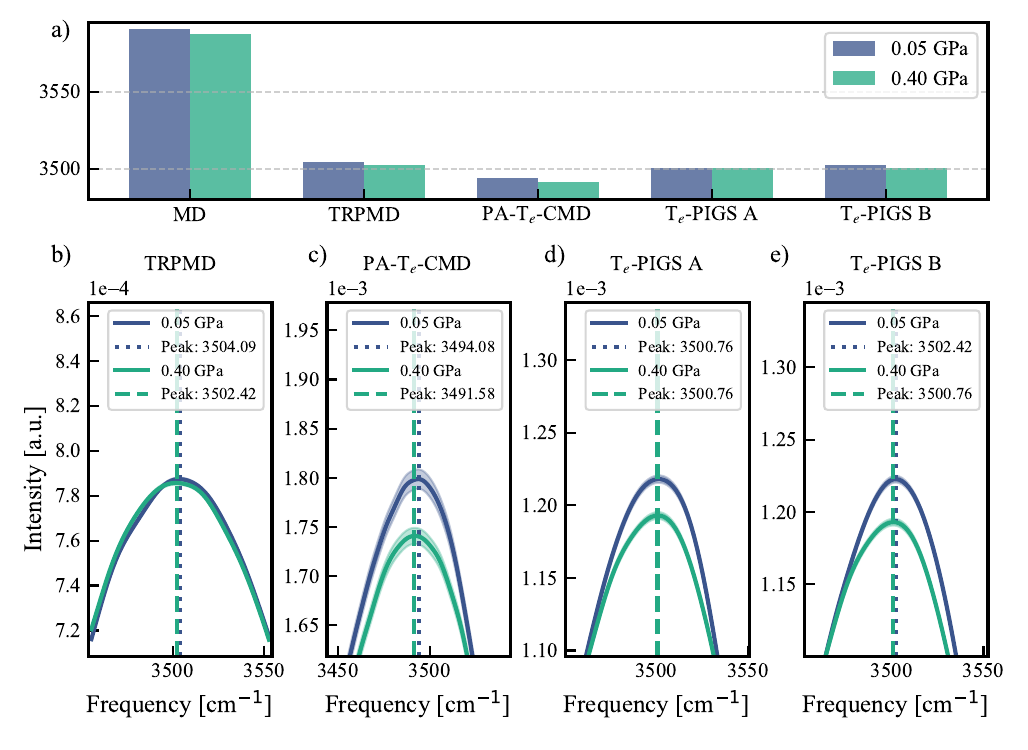}
\caption{Vibrational density of states (VDOS) of liquid water at different pressures, computed using classical MD, TRPMD, PA-T$_e$-CMD, and T$_e$-PIGS. The effects of nuclear quantum effects (NQEs) and pressure-induced spectral shifts are analyzed. The comparison highlights differences in peak resolution and broadening among the methods.}
\label{fig:S18}
\end{figure*}

Our VDOS results (Fig.~\ref{fig:S18}) serve to assess the capability of different path integral molecular dynamics (PIMD) methods in resolving these small frequency shifts. To determine the maximum of the O--H stretching band, we identify the frequency corresponding to the highest intensity in spectra smoothed using a Blackmann window \cite{Harris1978} with a 5 ps lag. Classical molecular dynamics (MD) simulations reproduce the expected trends: lower frequencies correspond to higher-pressure conditions. All PIMD methods exhibit a systematic red shift of approximately 85 cm$^{-1}$ relative to classical vibrational frequencies, reflecting nuclear quantum effects (NQEs) even at ultra-high pressures and room temperature.

Among the methods examined, thermostatted ring-polymer molecular dynamics (TRPMD) accurately reproduces the experimental trend despite suffering from spurious broadening, a known limitation of the method. Although small, the separation between vibrational peaks (approximately 4 cm$^{-1}$) is consistent between TRPMD and classical MD. Similarly, PA-T$_e$-CMD captures the correct experimental trend, with peak separations differing by less than 1 cm$^{-1}$ relative to TRPMD, yielding better resolved spectra.

Conversely, T$_e$-PIGS does not seem to resolve these subtle spectral shifts. This limitation likely stems from its reliance on a machine-learned potential of mean force (PMF), which may not fully capture pressure-dependent configurational changes. Specifically, when using a PMF trained only on the high-pressure system (denoted T$_e$-PIGS A), the method predicted nearly identical peak positions at both pressures, with differences within the spectral resolution. Only when a dedicated PMF is trained for each pressure condition (denoted T$_e$-PIGS B) does the method resolve the pressure-dependent spectral trend, with peak separations comparable to PA-T$_e$-CMD.

\begin{figure}[H]
    \centering
    \includegraphics[width=0.8\textwidth]{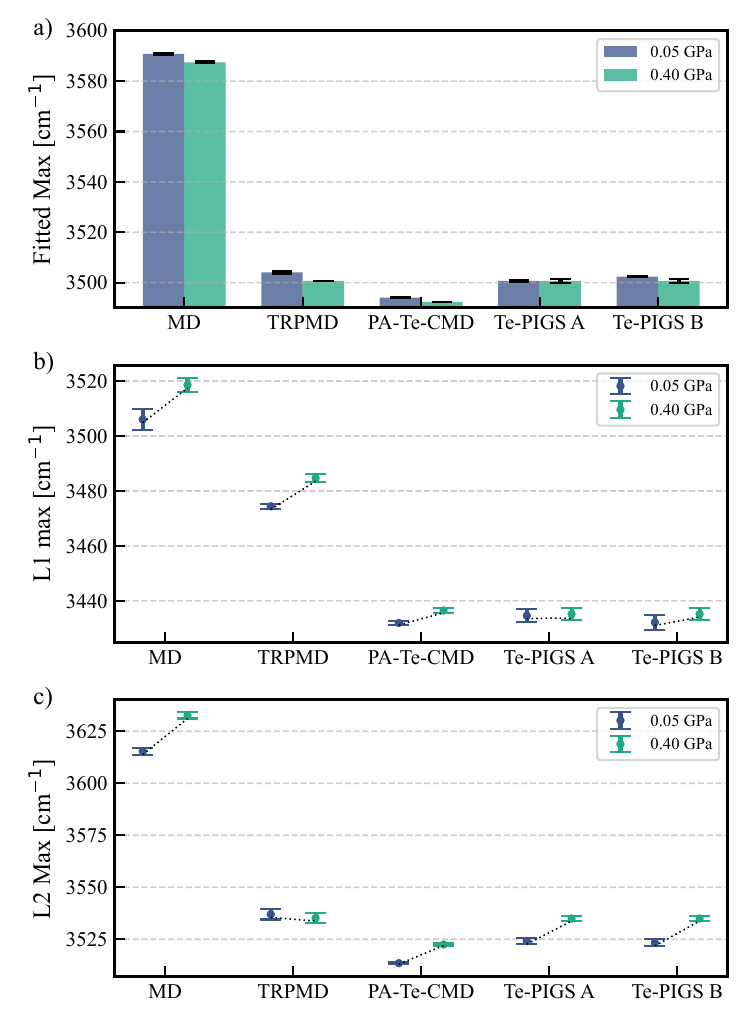} 
    \caption{(a) Bar chart comparing the fitted maximum frequencies obtained from a composite model for different simulation methods. (b) Error–bar plot of the first Lorentzian (L1) peak centers for the same methods and conditions. (c) Error–bar plot of the second Lorentzian (L2) peak centers. }
    
    \label{fig:S19}
\end{figure}

To further support our VDOS-based peak assignments, we performed a dynamically weighted nonlinear least-squares fitting of the O--H vibrational bands using two skewed Lorentzian functions to model water symmetric and asymmetric O--H stretching modes. The maxima of the fitted curves (Fig. \ref{fig:S19}a) confirm the same trend observed in the raw data (Fig.~\ref{fig:S18}). Moreover, the individual Lorentzian components L1 and L2 (Fig. \ref{fig:S19}b-c) show trends consistent with experimental high-pressure water studies \cite{Okada2005}, where both symmetric and asymmetric stretching frequencies shift to higher values with increasing pressure.


\newpage

\bibliography{bibliography}